\documentstyle[12pt,psfig]{article}
\makeatletter
\def\@cite#1#2{{[{#1}]\if@tempswa\typeout
{IJCGA warning: optional citation argument
ignored: `#2'} \fi}}

\newcount\@tempcntc
\def\@citex[#1]#2{\if@filesw\immediate\write\@auxout{\string\citation{#2}}\fi
 \@tempcnta\z@\@tempcntb\m@ne\def\@citea{}\@cite{\@for\@citeb:=#2\do
   {\@ifundefined
    {b@\@citeb}{\@citeo\@tempcntb\m@ne\@citea\def\@citea{,}{\bf ?}\@warning
     {Citation `\@citeb' on page \thepage \space undefined}}%
    {\setbox\z@\hbox{\global\@tempcntc0\csname b@\@citeb\endcsname\relax}%
    \ifnum\@tempcntc=\z@ \@citeo\@tempcntb\m@ne
   \@citea\def\@citea{,}\hbox{\csname b@\@citeb\endcsname}%
     \else
      \advance\@tempcntb\@ne
      \ifnum\@tempcntb=\@tempcntc
      \else\advance\@tempcntb\m@ne\@citeo
      \@tempcnta\@tempcntc\@tempcntb\@tempcntc\fi\fi}}\@citeo}{#1}}
\def\@citeo{\ifnum\@tempcnta>\@tempcntb\else\@citea\def\@citea{,}%
  \ifnum\@tempcnta=\@tempcntb\the\@tempcnta\else
   {\advance\@tempcnta\@ne\ifnum\@tempcnta=\@tempcntb \else
\def\@citea{--}\fi
   \advance\@tempcnta\m@ne\the\@tempcnta\@citea\the\@tempcntb}\fi\fi}
\makeatother

\def\boxit#1{\leavevmode\thinspace\hbox{\vrule\vtop{\vbox{\hrule%
        \vskip3pt\kern1pt\hbox{\vphantom{\bf/}\thinspace\thinspace%
        {\bf#1}\thinspace\thinspace}}\kern1pt\vskip3pt\hrule}\vrule}%
        \thinspace}
\def\Boxit#1{\noindent\vbox{\hrule\hbox{\vrule\kern3pt\vbox{
        \advance\hsize-7pt\vskip-\parskip\kern3pt\bf#1
        \hbox{\vrule height0pt depth\dp\strutbox width0pt}
        \kern3pt}\kern3pt\vrule}\hrule}}

\def\boxeq#1{\boxit{${\displaystyle #1 }$}}          

\newcommand{\Hh}{\lower1.2ex\hbox{$\stackrel{\textstyle
H}{\footnotesize\sim}$}}
\newcommand{\Hho}{\lower1.2ex\hbox{$\stackrel{\textstyle
H_1}{\footnotesize\sim}$}}
\newcommand{\Hhw}{\lower1.2ex\hbox{$\stackrel{\textstyle
H_2}{\footnotesize\sim}$}}
\newcommand{\h}{\lower1.2ex\hbox{$\stackrel{\textstyle
h}{\footnotesize\sim}$}}

\newcommand{\nn}{\nonumber}

\newcommand{\gsim}{\lower.7ex\hbox{$\;\stackrel{\textstyle>}{\sim}\;$}}
\newcommand{\lsim}{\lower.7ex\hbox{$\;\stackrel{\textstyle<}{\sim}\;$}}
\newcommand{\be}{\begin{equation}}
\newcommand{\ee}{\end{equation}}
\newcommand{\bea}{\begin{eqnarray}}
\newcommand{\eea}{\end{eqnarray}}
\def\bma#1{\mbox{\boldmath{$#1$}}}
\def\simlt{\stackrel{<}{{}_\sim}}

\def\baselinestretch{1}
\begin{document}
\catcode`@=11
\newtoks\@stequation
\def\subequations{\refstepcounter{equation}%
\edef\@savedequation{\the\c@equation}%
  \@stequation=\expandafter{\theequation}
  \edef\@savedtheequation{\the\@stequation}
  \edef\oldtheequation{\theequation}%
  \setcounter{equation}{0}%
  \def\theequation{\oldtheequation\alph{equation}}}
\def\endsubequations{\setcounter{equation}{\@savedequation}%
  \@stequation=\expandafter{\@savedtheequation}%
  \edef\theequation{\the\@stequation}\global\@ignoretrue

\noindent}
\catcode`@=12
\begin{titlepage}

\title{{\bf 
Unconventional low-energy SUSY
from warped geometry}}
\vskip2in
\author{  {\bf J.A. Casas$^{1,2}$\footnote{\baselineskip=16pt E-mail: {\tt
alberto@makoki.iem.csic.es}}},
{\bf J.R. Espinosa$^{2,3}$\footnote{\baselineskip=16pt E-mail: {\tt
espinosa@makoki.iem.csic.es}}} and 
{\bf I. Navarro$^{1}$\footnote{\baselineskip=16pt E-mail: {\tt
ignacio@makoki.iem.csic.es}}}
\hspace{3cm}\\
 $^{1}$~{\small I.E.M. (CSIC), Serrano 123, 28006 Madrid, Spain}.
\hspace{0.3cm}\\
 $^{2}$~{\small I.F.T. C-XVI, U.A.M., 28049 Madrid, Spain}
\hspace{0.3cm}\\
 $^{3}$~{\small I.M.A.F.F. (CSIC), Serrano 113 bis, 28006 Madrid, Spain}
} 
\date{} 
\maketitle 
\def\baselinestretch{1.15} 
\begin{abstract}
\noindent 
Supersymmetric models with a warped fifth spatial dimension can solve
the hierarchy problem, avoiding some shortcomings of
non-supersymmetric  constructions, and predict a plethora of new phenomena
at typical scales $\Lambda$ not far from the electroweak scale
($\Lambda\sim$ a few TeV).  In this paper we derive the
low-energy effective theories of these models, valid at energies below
$\Lambda$.  We find that, in general, such effective
theories can deviate significantly from the Minimal Supersymmetric
Standard Model (MSSM)  or other popular extensions of it, like the
NMSSM: they have non-minimal  K\"ahler potentials (even in the 
$M_p\rightarrow \infty$ limit), 
and the radion is coupled to the visible
fields, both in the superpotential and the K\"ahler potential, in a
non-trivial (and quite model-independent) fashion.  The corresponding
phenomenology is pretty unconventional, in particular the electroweak
breaking occurs in a non-radiative way, with  $\tan\beta\simeq 1$ as a
quite robust prediction, while the mass of the lightest Higgs boson
can be as high as $\sim$ 700~GeV.
\end{abstract}

\thispagestyle{empty}
\vspace*{3cm}
\leftline{September 2001}
\leftline{}

\vskip-22cm
\rightline{}
\rightline{IFT-UAM/CSIC-01-24}
\rightline{IEM-FT-218/01}
\rightline{hep-ph/0109127}
\vskip3in

\end{titlepage}
\setcounter{footnote}{0} \setcounter{page}{1}
\newpage
\baselineskip=20pt

\noindent

\section{Introduction}
\setcounter{equation}{0}
\renewcommand{\theequation}{1.\arabic{equation}}

In the last two years there has been a lot of interest in warped
compactifications \cite{RS,warped} as an alternative mechanism to
address the hierarchy problem.  In its simplest realization, the
Randall-Sundrum (RS) model \cite{RS}, the action is 5D gravity realized on 
${\cal M}_4\times
S_1/Z_2$ with negative cosmological constant
and two constant three-brane energy-densities located at the
$Z_2$ fixed points.
For an appropriate tuning of the latter
with the bulk cosmological constant, 
the metric admits the RS
solution\footnote{We use the metric signature $(+ - - - -)$.}
\be 
ds^2=e^{-2k|x^5|}\eta_{\mu\nu} dx^\mu dx^\nu - (dx^5)^2\;.
\label{RS}
\ee
Here $x^5\equiv \vartheta r_0$, where the angle $\vartheta$ 
($-\pi\leq \vartheta\leq \pi$) 
parametrizes the extra dimension, with $\vartheta$ and $-\vartheta$
identified; $r_0$ is a free
parameter giving the radius of the compact $S_1$ and $k={\cal O}(M_5)$, with
$M_5$ the 5D Planck mass (the 4D Planck mass is $M_p^2\simeq M_5^3/k$). 
The branes are located at $x^5=0$ 
(``hidden brane'') and $x^5=\pi r_0$ (``visible brane''). In the
original RS model, all the visible fields (except gravity) are located at
the visible brane, and this will be also our assumption here.
Then, any mass parameter, $m_0$, at the visible brane 
gets effectively suppressed by the warp factor giving
$m = e^{-kr_0\pi} m_0$ \cite{RS}. For moderate values of $r_0$,  
the mass
hierarchy achieved is of the right order, so that $m\sim\cal O$~(TeV) 
for the natural choice  $m_0\sim M_p$.
This is the simplest and probably most attractive scenario, although
there are variations in the literature.

At first sight, it may seem unnecessary to supersymmetrize these 
warped constructions. After all, low-energy supersymmetry (SUSY) is an
alternative extension of the Standard Model whose primary motivation
is also to face the hierarchy problem. However, SUSY may prove very helpful to avoid
some shortcomings of the original RS scenario. In particular, it can provide
an explanation (at least a partial one) for the correlations between the
brane tensions and the bulk cosmological constant \cite{correlation}.
Moreover, SUSY may help to protect the hierarchy against destabilization by radiative 
corrections. Finally, SUSY is likely a necessary ingredient to make contact 
with warped superstring constructions.
As a result, a lot of effort has been dedicated 
to supersymmetrize warped scenarios \cite{correlation,LS,BNZ,Pok,SUSYwarp}. 
The stabilization 
of these constructions, and in particular the value of $r_0$, is generically
connected to the mechanism of SUSY breaking. 

Still, from the point of view of low-energy SUSY, one may wonder what
is gained by promoting any ordinary version of the MSSM to a higher
dimensional warped world. A partial answer is that the warped version 
has a solution to the $\mu$-problem built in. As any other mass
scale in the visible brane, the value of $\mu$ is affected by the warp
suppression factor, so that $\mu$ is naturally $\cal O$~(TeV). However,
a more complete answer to this question requires the phenomenology of 
the supersymmetric warped worlds to be  worked out.

This is the main goal of this paper: to derive the low-energy 
effective theories of supersymmetric warped constructions and to extract from 
their the low-energy phenomenology. We will show that, indeed, these
effective theories (and the phenomenology they imply) are  
non-conventional.  This might be useful to alleviate
some of the drawbacks of the MSSM. We will show in
particular that the upper bound on the Higgs mass, which is
starting to be worrisome in the conventional MSSM, evaporates. 
On the  other hand, we will see that  a
non-conventional phenomenology provides many useful tests to
distinguish a warped SUSY world from an ordinary MSSM (or further
extensions, as the NMSSM). Conversely, the present experimental
information imposes non-trivial constraints on supersymmetric warped
constructions.

In section 2 we briefly review the construction of SUSY warped
scenarios and the extraction of the 4D theory by matching to the 5D
complete theory.  In the process, we fix some notation and make
explicit our starting assumptions (in particular 
for the choice of K\"ahler potential). In section~3 we derive the
low-energy effective theory in the $M_p\rightarrow\infty$ limit,
obtaining the globally-supersymmetric effective theory, which turns out 
to be non-minimal. We
also make some preliminary  analysis of the differences that arise
with respect to the MSSM.  The problem of supersymmetry breaking and
its effects on the low-energy theory, in particular on the generation of
soft terms, are discussed in section~4. Armed with the complete
effective theory (which now includes also supersymmetry breaking
effects) we examine some phenomenological  implications in later
sections. Electroweak symmetry breaking is the subject of section~5,
while section~6 is devoted to the analysis of the Higgs sector
and section~7 to the neutralino sector.
Finally, section~8 contains a summary and some conclusions. Appendix~A gives the
Lagrangian for a globally-supersymmetric theory with general K\"ahler potential
and gauge kinetic function. Appendix~B examines more briefly an alternative
choice of K\"ahler potential (different from the one adopted in the main text).  
Appendix~C contains the technical details of the calculation of the 
Higgs spectrum.

\section{Supersymmetric warped scenarios}
\setcounter{equation}{0}
\renewcommand{\theequation}{2.\arabic{equation}}

The starting point for supersymmetric warped constructions is 
the 5D SUGRA lagrangian density
\bea
\label{sugra5}
{\cal L}_{\rm SUGRA,5} = &M_5^3 & \biggl\{ \sqrt{G} \left[
\frac{1}{2} R(G)
- \frac{1}{4} C^{MN} C_{MN} + 6 k^2 \right]
\nonumber\\
&-& \frac{1}{6\sqrt{6}} \epsilon^{MNPQR} B_M C_{NP} C_{QR}
+ \hbox{\rm fermion\ terms} \biggr\},
\eea
where $M_5$ is the 5D Planck mass; 
$M, N, \ldots = 0, \ldots, 3, 5,$ are 5-dimensional
spacetime indices; $G_{MN}$ is the 5-dimensional metric 
(with $G={\rm Det}[G_{MN}]$);
$-6k^2M_5^3$ is the bulk cosmological 
constant and
$C_{MN} = \partial_M B_N - \partial_N B_M$
is the field strength of the graviphoton, $B_M$.
This theory is realized on ${\cal M}_4\times
S_1/Z_2$ with appropriate $Z_2$ assignment of charges. In addition,
one has to include the pieces corresponding to the two energy-densities 
located at the two $Z_2$ fixed points
\bea
\label{boundary}
\Delta{\cal L}_5 = - \delta(x^5) \sqrt{-g_1}\, V_1
- \delta(x^5 - \pi r_0) \sqrt{-g_2}\, V_2,
\eea
where $g_{1,2}$ are the induced 4-dimensional metrics on the boundaries
and $V_{1,2}$ are constants.
The initial 5D $N=2$ SUGRA of eq.~(\ref{sugra5}) is broken down to 4D $N=1$
SUGRA by the boundary terms (\ref{boundary}) and the  $Z_2$ orbifold
projection.  Then, provided that ${V_1} = -{V_2} = {6 M_5^3}k$,
 the theory admits the RS
solution (\ref{RS}) with $r_0$ and $b_0\equiv
B_0$ arbitrary, and $B_\mu=0$. 
The massless 4D fluctuations of these parameters, plus
those of the 4D metric, correspond to the replacements $r_0\rightarrow
r(x)$, $b_0\rightarrow b(x)$ and  $\eta_{\mu\nu}\rightarrow
g_{\mu\nu}(x)$.  In fact, the corresponding metric
\be 
ds^2=e^{-2kr(x)|\vartheta|}g_{\mu\nu}(x) dx^\mu dx^\nu - r^2(x)
(d\vartheta)^2\;,
\label{RS2}
\ee
is shown \cite{LS} to be a solution of the 5D action (up to higher
derivative terms).

The lagrangian density  ${\cal L}_{\rm SUGRA,5} +\Delta{\cal L}_5$
must be supplemented with extra pieces in order to stabilize the
radion field, $r$ (and $b$), at some vacuum expectation values
(VEVs). This is absolutely necessary for several reasons. In the
absence of a potential, the radion  couples to 4D gravity as a
Brans-Dicke scalar with a strength incompatible with precision tests
of general relativity.  In addition, light weakly-coupled moduli are
known to produce cosmological problems (the mass required to avoid
them depends on  the size of the couplings). The radion  stabilization
is intimately connected to the mechanism of SUSY breaking since, if
SUSY is unbroken, the radion potential is flat~\cite{Pok}. Several
models have been proposed  for radion stabilization in this context.
In ref.~\cite{LS} this is achieved by the inclusion of two
super-Yang-Mills (SYM) sectors, one in the bulk and one in the hidden
brane, plus a SUSY breaking sector also located in the hidden
brane. In ref.~\cite{Pok}  stabilization is achieved by means of source
superpotentials (essentially constants) located at the two branes,
coupled to scalar fields living in the bulk. Actually, in
ref.~\cite{Pok} the bulk is populated not only by the 5D gravitational
sector [eq.~(\ref{sugra5})], but also by  a number of (neutral) bulk
hypermultiplets. This is reasonable if the whole picture is to be
embedded in a string compactification, where moduli fields (including
the universal dilaton field) abound.   Finally, one has to introduce
matter and gauge fields in the visible brane.  In the effective 4D
SUGRA theory the $r$ and $b$ fields combine in a complex scalar field
$T = k\pi(r + ib\sqrt{2/3})$  (this can be readily checked by noting
that the bulk SYM action upon integration  in the fifth dimension
contains the pieces  $\sim r\ {\rm tr} FF +  b\ {\rm tr} \tilde FF$
with the right coefficients \cite{LS,BNZ,Pok}).

Following refs.~\cite{LS,BNZ}, the $T$-dependent part of the K\"ahler
potential in 4D SUGRA, $K(T,T^\dagger)$, can be obtained by comparing
the curvature term of the action (\ref{sugra5}) upon integration in
the compact dimension \cite{RS}
\bea
\label{leff4}
S_{\rm 4,eff} =  \frac{1}{2}{M_p^2} \int d^4 x\, \sqrt{-g} \left[
\left( 1 - e^{-2 \pi k r(x)} \right) R(g) + \cdots \right]\ , \eea
where $M_p^2\equiv M_5^3/k$, with the general form of a 4D SUGRA
potential
\bea
\label{SUGRA4app}
{S}_{\rm SUGRA,4}= \int d^4 x d^4\theta\, E \phi_c^\dagger \phi_c
\Phi(T, T^\dagger) = -\int d^4 x\sqrt{-{g}} \biggl[ \frac{1}{6} \Phi
R({g}) + \cdots \biggr].  \eea
Here $E$ is the superspace determinant of the vierbein superfield,
$\phi_c$ is the conformal compensator and $\Phi(T, T^\dagger)$ is
real.  The right-hand side of  eq.~(\ref{SUGRA4app}) is written in the
minimal Poincar\'e  superconformal gauge, $\phi_c = 1 +\theta^2
F_c$. Before doing this fixing, the left hand side of
eq.~(\ref{SUGRA4app}) is formally Weyl-invariant, so one can assume
that it is written in the same conformal frame as eq.~(\ref{leff4}),
i.e. the 4D metric is the same in both. Thus \cite{LS,BNZ,Pok}
\be
\label{PhiTT0}
\Phi(T,T^\dagger)=  3M_p^2 \left[ \exp[- (T + T^\dagger)] - 1 \right]\ ,
\ee
and
\be K(T,T^\dagger)= -3M_p^2\log\left[-{1\over
3M_p^2}\Phi(T,T^\dagger)\right]\ .
\label{PhiTT}
\ee

In addition, one has to consider the pieces coming from the two
branes.  Since the hidden brane corresponds to $\vartheta=0$, the
metric induced  by eq.~(\ref{RS2}) in the hidden brane and ${\cal
L}_{\rm 4D, hid}$ are  independent of $r(x)$. This implies
$T$-independence of the supersymmetric  theory in the hidden brane
\cite{LS}. Consequently, in this conformal  frame ${\cal L}_{\rm 4D,
hid}$ must take the form \cite{LS}
\be
\label{Lhidapp}
{\cal L}_{\rm 4D,hid}= -\frac{1}{2}\biggl[\phi_c^\dagger \phi_c
\Phi_{\rm hid}(\hat \varphi_{\rm hid}, \hat \varphi_{\rm
hid}^\dagger)\biggr]_D + \biggl[ f_{\rm hid}(\hat \varphi_{\rm hid})
{\rm tr}(W'^\alpha W'_\alpha)\biggr]_F + \biggl[\phi_c^3 W_{\rm
hid}(\hat \varphi_{\rm hid}) \biggr]_F\ .  \ee
Here $\hat \varphi_{\rm hid}$, $W'^\alpha$ are the matter and  the
gauge field-strength chiral superfields confined to the hidden brane;
$\Phi_{\rm hid}$ is real, and $f_{\rm hid}$, $W_{\rm hid}$  (the gauge
kinetic function and the superpotential) are holomorphic.  These
functions can also depend on the (even component under the
$Z_2$-parity of the) bulk hypermultiplets to be introduced below.

For the visible brane it is convenient to first perform a Weyl
rescaling of the metric $g_{\mu\nu}\rightarrow
e^{2kr(x)\pi}g_{\mu\nu}$. In this frame the induced metric in the
visible brane is independent of $r$, thus the form of ${\cal L}_{\rm
4D, vis}$ becomes completely analogous to (\ref{Lhidapp}). This
rescaling is equivalent to a redefinition of the conformal compensator
$\phi_c\rightarrow e^{T}\phi_c$,  prior to fixing the superconformal
gauge in the usual way. Therefore, in terms of the original metric,
${\cal L}_{\rm 4D, vis}$ must read
\bea
\label{Lvis}
{\cal L}_{\rm 4D,vis}&=&-\frac{1}{2} \biggl[\phi_c^\dagger \phi_c
e^{-(T+T^\dagger)} \Phi_{\rm vis}(\hat \varphi_{\rm vis}, \hat
\varphi_{\rm vis}^\dagger)\biggr]_D\nonumber\\ &+& \biggl[ f_{\rm
vis}(\hat \varphi_{\rm vis}) {\rm tr}(W'^\alpha
W'_\alpha)\biggr]_F+\biggl[ \phi_c^3 e^{-3T} W_{\rm vis}(\hat
\varphi_{\rm vis}) \biggr]_F \ .  \eea
{}From eqs.~(\ref{PhiTT}--\ref{Lvis}) one gets  the complete K\"ahler
potential $K=-3M_p^2\log[-\Phi/(3M_p^2)]$, where
$\Phi=\Phi(T,T^\dagger) + \Phi_{\rm hid}
+e^{-(T+T^\dagger)}\Phi_{\rm vis}
$. This already shows the most distinctive features of 4D SUSY arising
from warped constructions: the radion, $T$, is coupled to the visible
fields, {\em both} in the superpotential and the K\"ahler potential,
in a non-trivial (and quite model-independent) fashion. As we will
see, this results in a new and characteristic
phenomenology.

Eqs.~(\ref{PhiTT}--\ref{Lvis}) were first obtained in ref.~\cite{LS},
where the possible presence of additional moduli  in the bulk (which
seems likely, as mentioned before) was not considered. This was done
in ref.~\cite{Pok} by representing such matter as bulk
hypermultiplets.  In particular, ref.~\cite{Pok} considered the presence of a
``universal hypermultiplet'' which (with a suitable choice  of
quaternionic sigma-model metric and couplings to the branes)
reproduces in 4D the basic features of the usual dilatonic multiplet,
$S$,  of string theories:  it contributes with a term $-M_p^2\log(S +
S^\dagger)$ to the K\"ahler potential and it is coupled to the gauge
multiplets  in both branes in the universal way characteristic of the
dilaton,  i.e.  $f_{\rm hid}=f_{\rm vis}=S$.

In general, we indeed expect that, in first approximation, the
contribution of the hypermultiplets to the K\"ahler potential amounts
to a separate term, say $K(S,S^\dagger)$. This is because, in the
frame where  eqs.(\ref{leff4}, \ref{SUGRA4app}) are written, the
kinetic term of $S$ gets a factor $\Phi(T,T^\dagger)$, which simply
arises from the integration in the extra dimension. Then, after a
conformal redefinition to leave the curvature term in the canonical
form, $\frac{1}{2}R$ (as must be done to write the SUGRA Lagrangian in
the usual fashion of Cremmer et al. \cite{Cremmer,WessBagger})  the
$S$-kinetic term becomes $T$-independent.  Thus, the latter must arise from an
independent piece in the K\"ahler potential.  On the other hand, it is
known that in string theories the dilaton K\"ahler potential may
acquire sizeable perturbative and non-perturbative  contributions
\cite{kahlercorr}, which can modify its form  in a substantial
manner. Therefore, we prefer to leave the dilaton part of the K\"ahler
potential, $K(S,S^\dagger)$, as an unknown function, although we will
use $K(S,S^\dagger)=-M_p^2\log(S + S^\dagger)$ in some computations with
no loss of generality. With respect to the gauge kinetic functions,
$f$, there can be important corrections to the previous universal
assumption. Actually, the universal condition is not desirable in this
context since it implies unification of observable gauge couplings at
the TeV scale, which is phenomenologically untenable. On the other
hand, non-universality of gauge couplings  may arise in actual string 
constructions for
several reasons. This happens in the context of the weakly-coupled
heterotic string for different Kac-Moody levels of the various gauge
groups and/or due to sizeable moduli-dependent corrections. In the
context of Type I string constructions, the $f$-function for a given
$p$-brane consists of a universal part (which is either the dilaton or
a modulus) plus a model- and gauge-group-dependent sizeable
contribution proportional to the twisted moduli. In consequence, we
will not make strong assumptions about the form of $f_{\rm hid},f_{\rm
vis}$. Our only assumption is that $f_{\rm vis}$ is  independent of
$T$ (which is true if the observable gauge bosons live in the visible
brane). In our calculations  the dilaton will only play a role in
breaking supersymmetry,  since this breaking will occur along the
$F_S$-direction. However, the details   about the explicit form of
$K(S,S^\dagger)$ and $f_{\rm vis}$ turn out  to be unimportant. Actually, in 
our set-up $S$ may represent
the ordinary dilaton or any other moduli field living in the bulk. In
summary, our starting point for the  effective 4D 
superpotential is
\bea
\label{EffW}
W=W_{\rm hid} + e^{-3T} W_{\rm vis}\ , 
\eea
and for the K\"ahler potential\footnote{There is a misprint in formula (3.17)
of \cite{LS} which does not have  the $\exp[-(T+T^\dagger)]$ factor in
front of $\Phi_{\rm vis}$.}
\bea
\label{EffK}
K= K(S,S^\dagger) - 3M_p^2\log\biggl[\frac{\Phi(T,T^\dagger) +
\Phi_{\rm hid}(\hat \varphi_{\rm hid}, \hat \varphi_{\rm hid}^\dagger)
+e^{-(T+T^\dagger)}\Phi_{\rm vis} (\hat \varphi_{\rm vis}, \hat
\varphi_{\rm vis}^\dagger)} {-3M_p^2}\biggr] 
\eea
where $\Phi(T,T^\dagger)$ is given by eq.~(\ref{PhiTT}).  For later
use we define $\Sigma_{\rm vis}(\hat \varphi_{\rm vis})$ through
\bea
\label{Sigmavis}
\Phi_{\rm vis}\equiv 3M_p^2 \left\{\exp\left[\Sigma_{\rm
vis}/(3M_p^2)\right]  - 1 \right\}\ , \eea
so that eq.~(\ref{EffK}) takes the form
\bea
\label{EffK2}
K= K(S,S^\dagger) - 3M_p^2\log\biggl[ 1-\exp\left\{- (T + T^\dagger)
+\frac{\Sigma_{\rm vis}(\hat \varphi_{\rm vis})}{3M_p^2} \right\}-
{\Phi_{\rm hid}(\hat \varphi_{\rm hid})\over 3M_p^2} \biggr]\ .  \eea

One may wonder about the explicit form of $\Sigma_{\rm vis}$ in
eq.~(\ref{EffK2}) [or, equivalently, $\Phi_{\rm vis}$ in
eq.~(\ref{EffK})].  From the point of view of 4D SUGRA it  is not
possible to theoretically constrain its form.  On the other hand, in
ref.~\cite{Pok} it is argued that, very likely, the actual 4D SUGRA
actions from 5D  are just a subset of all possible 4D SUGRA actions.
In this sense, the authors of ref.~\cite{Pok}  give an ansatz for the
visible fields, consistent with the dimensional reduction from
5D, which corresponds to $\Sigma_{\rm vis} = \sum_i |\hat
\varphi_i|^2$.  
Intuitively, a justification for this is the
following \cite{Pok}. If we start with canonical kinetic  terms [except for the
$e^{-(T+T^\dagger)}$ scaling] for the  $\hat \varphi_{\rm vis}$ fields
in the initial conformal frame  [in which eqs.~(\ref{leff4},
\ref{SUGRA4app}) are written], then, in the limit of small $k$, it is
easy to see that eq.~({\ref{EffK}) must render $K= K(S,S^\dagger) -
3M_p^2\log\left[T+T^\dagger - |\hat \varphi_{\rm
vis}|^2/(3M_p^2)\right]$.  This suggests that in general the K\"ahler
potential will depend on the $T$ and $\hat \varphi_{\rm vis}$ fields
through the  combination $T+T^\dagger -|\hat \varphi_{\rm
vis}|^2/(3M_p^2)$,  and this leads to $\Sigma_{\rm vis} = \sum_i |\hat
\varphi_i|^2$.

Throughout the paper we will leave $\Sigma_{\rm vis}$ as an unknown
function for the general derivations, but for concrete computations we
will follow ref.~\cite{Pok} and use the above ansatz.  On the other
hand, besides the $\sim |\hat \varphi|^2$ terms in  $\Sigma_{\rm
vis}$, there may be additional ``chiral'' terms, like  $\sim
(\hat H_1\cdot \hat H_2 + {\mathrm h.c.})$ for the Higgses
[here $\cdot$ stands for the $SU(2)$ product: $\hat H_1\cdot \hat
H_2\equiv i  \hat H_1^T{\bma \sigma}_2 \hat H_2$]. Such terms are
relevant for  phenomenology and we will allow their presence. Hence,
we will take
\bea
\label{ansatzSigma}
\Sigma_{\rm vis} = \sum_i |\hat \varphi_i|^2 + \left(\lambda \hat
H_1\cdot \hat H_2 + {\mathrm h.c.}\right)\ .  \eea

For other ans\"atze, most of the basic features of the
non-conventional SUSY phenomenology associated to these scenarios
(which will be analyzed  in the following sections) still hold. As
stated after eq.(\ref{Lvis}), this is a consequence of the non-trivial
and quite model-independent way in which the radion is coupled to the
observable fields in $K$ and $W$. This holds in particular  if one
takes the somewhat simpler ansatz  $\Phi_{\rm vis} = \sum_i |\hat
\varphi_i|^2$.  This case is briefly discussed in Appendix~B.

With regard to the dependence of $W_{\rm vis}$ on the observable
fields,  we will consider a MSSM-like superpotential, with the
ordinary Yukawa terms  plus a $\mu$ mass term for the Higgses
(possible non-renormalizable terms will be   discussed later on):
\bea
\label{Wvis}
W_{\rm vis}(\hat \varphi_{\rm vis}) =\hat  h_{U}^{ij} \hat Q_{Li}\cdot
\hat H_2 \hat U_{Rj} +\hat h_{D}^{ij} \hat H_1\cdot \hat Q_{Li}  \hat
D_{Rj} + \hat \mu_0 \hat H_1\cdot \hat H_2 \ .  \eea
where $i,j=1,2,3$ are family indices.  The value of $\hat \mu_0$ is
naturally of the order of the fundamental scale of the theory, i.e.
$\hat \mu_0 = {\cal O}(M_p)$.

\section{Effective supersymmetric theory}
\setcounter{equation}{0}
\renewcommand{\theequation}{3.\arabic{equation}}

In this section we derive the effective globally-supersymmetric
theory  at low-energies that arises from the  4D superpotential and
K\"ahler potential displayed in section~2,
eqs.~(\ref{EffK}--\ref{Wvis}).  The contributions from SUSY breaking
will be studied later, in sect.~4.

In order to extract the effective theory, we have first to take into
account that the radion field, $T$, takes an expectation value such
that
\be  e^{-\langle T+T^\dagger\rangle}\equiv{\Lambda^2\over 3 M_{p}^2}\ ,
\label{Tvev}
\ee
where $e^{-\langle T+T^\dagger\rangle}$ is the usual warp suppression
factor and ${\Lambda}={\cal{O}}$(TeV) is essentially the effective
fundamental  mass-scale in the visible brane.  This means, in
particular, that $T$ and the visible fields $\hat\varphi_i$ (we will
remove the subscript ``vis''  if  there is no risk of confusion)  in
eqs.~(\ref{EffK}--\ref{Wvis})  are far from being correctly
normalized. We define $t$, $\varphi_i$ as fields with 
canonical kinetic terms (before electroweak
breaking) by
\bea
\label{normal}
T = \langle T \rangle + t/\Lambda\ ,\;\;\;  \varphi_i =
\frac{\Lambda}{\sqrt{3}M_p} \hat\varphi_i\ .  \eea
Then, the effective K\"ahler potential at low energy is obtained from
eq.~(\ref{EffK2}) by taking the limit $M_p\rightarrow\infty$ (with
$\Lambda$ fixed).  Focusing just on the observable sector, this gives
\be \boxeq{ K_{\rm eff}=\Lambda^2 \exp\left\{-{t+t^*\over\Lambda}+
{1\over\Lambda^2}\Sigma_{\rm eff}(\varphi_i) \right\} }
\label{Keff}
\ee
where $\Sigma_{\rm eff} \equiv(\Lambda^2/ 3 M_{p}^2) \Sigma_{\rm vis}$.
Since it is quadratic in the fields, $\Sigma_{\rm eff}$ has the same
form as $\Sigma_{\rm vis}$, eq.~(\ref{ansatzSigma}),  in terms of the
canonically normalized fields:
\bea  \Sigma_{\rm eff} = \sum_i | \varphi_i|^2 + \left(\lambda
H_1\cdot  H_2 + {\mathrm h.c.}\right)\ .
\label{Seff}
\eea
The low energy effective superpotential in the observable sector is
given by \cite{Weff} $W_{\rm eff}=e^{\langle K/2M_p^2\rangle}
e^{-3T}W_{\rm vis}$.  So, using eqs.~(\ref{Wvis}, \ref{Tvev},
\ref{normal}), we get
\be \boxeq{ W_{\rm eff}=e^{-3t/\Lambda}\left[ h_{U}^{ij} Q_{Li}\cdot
H_2  U_{Rj} + h_{D}^{ij} H_1\cdot Q_{Li}  D_{Rj} + \mu_0 H_1\cdot H_2
\right] }
\label{Weff}
\ee
with
\bea  h_{U,D}^{ij} & = &e^{\langle K/2M_p^2\rangle}  e^{-3i\langle
{\rm Im}T\rangle} \hat h_{U,D}^{ij} ={\cal O}(1) \hat h_{U,D}^{ij}\ ,
\nonumber\\ 
\mu_0 & = &e^{\langle K/2M_p^2\rangle}  e^{-3i\langle {\rm
Im}T\rangle} \left(\frac{\Lambda}{\sqrt{3}M_p}\right) \hat \mu_0
={\cal O}(\Lambda) \ ,
\label{redef}
\eea
where $e^{\langle K/2M_p^2\rangle}  \simeq {\cal O}(1)$.  Equation
(\ref{redef}) illustrates the fact that, once the fields are properly
normalized, all the mass scales in the observable sector (in this case the
$\mu$ term) get a suppression factor ${\cal O}(\Lambda/M_p)$. This in
particular means that there is no $\mu$-problem in this context, since
$\mu_0$ has naturally the right order of magnitude.

$W_{\rm eff}$ and $K_{\rm eff}$ in (\ref{Weff}) and (\ref{Keff})
describe a globally-supersymmetric theory valid at energies below the
scale $\Lambda$, which plays the role of UV cut-off\footnote{Non-minimal 
K\"ahler  potentials that result from 
decoupling  heavy fields are discussed in \cite{decoup}.}.
Physically, $\Lambda$ corresponds roughly to the mass scale of the lightest
Kaluza-Klein excitations\footnote{More precisely, the mass of the
lightest K-K excitations is \cite{RS,LS} $m_{KK}\simeq \frac{\Lambda
k}{\sqrt{3}M_p}=
\frac{\Lambda}{\sqrt{3}}\left(\frac{k}{M_5}\right)^{3/2}$.}.  The
supersymmetric Lagrangian (before SUSY breaking) that follows from
(\ref{Weff}) and (\ref{Keff})
can be obtained in terms of component fields  using the general
formulae  for non-minimal K\"ahler potentials given in Appendix~A.

If the scale $\Lambda$ were much higher than  $\mu_0$, this scenario
would reproduce the MSSM: an expansion in powers of $\mu_0/\Lambda$
and $\varphi_i/\Lambda$ gives
\be  W_{\rm eff}\simeq  h_{U}^{ij} Q_{Li}\cdot H_2  U_{Rj} +
h_{D}^{ij} H_1\cdot  Q_{Li}  D_{Rj} + \mu_0 H_1\cdot H_2 +... \equiv
W_{\rm MSSM}+...\ ,
\label{Weffl}
\ee
and
\be K_{\rm eff}\simeq \Lambda^2 - \Lambda (t+t^*)+{1\over 2}
(t+t^*)^2+|H_1|^2+|H_2|^2+\left( \lambda H_1\cdot H_2 + {\mathrm
h.c.}\right)+...
\label{Keffl}
\ee
with terms suppressed by inverse powers of $\Lambda$ omitted.
Throwing away terms that give a zero contribution to the $D$-term of
$K_{\rm eff}$ (and so, do not contribute to the Lagrangian), the
K\"ahler potential in (\ref{Keffl}) is equivalent  to the simpler
minimal one
\be
\label{Kmin}
K_{\rm eff}\simeq |t|^2 +|H_1|^2+|H_2|^2+...\equiv K_{\rm min}+...  \ee
In this limit, therefore, the Lagrangian reduces  to that of the MSSM
(the radion field, $t$, simply decouples from the Standard fields).

We are, of course, interested in a different regime  with
$\Lambda\sim$ 4-10 TeV. The lower limit\footnote{The precise value for
this lower limit depends on the value of $k/M_p$, see~\cite{explim}.
The quoted value corresponds to $k/M_p\sim 0.5$.} is roughly what 
is required to
satisfy experimental limits on exotic KK excitations  of gravitational
fields \cite{explim}.  The upper limit is set by naturalness criteria,
as we wish to avoid reintroducing a hierarchy problem. Therefore,
$\Lambda$ should not be much larger than $\mu_0$ (or than the
gravitino mass, $m_{3/2}$, as is discussed in the next section).  In
this case, $K_{\rm eff}$ and $W_{\rm eff}$, as given in
eqs.~(\ref{Keff}, \ref{Weff}), do not reduce to the minimal ones,
eqs.~(\ref{Weffl}, \ref{Kmin}), so  we expect important deviations
from ordinary supersymmetric versions  of the SM, like the MSSM or the
NMSSM \cite{nmssm}. Such deviations are due to
\begin{description}

\item[i)] The presence of new light degrees of freedom,  associated to
the radion superfield, $t$.

\item[ii)] Mixing of $t$ with the Higgs fields through kinetic and
mass terms after electroweak symmetry breaking. A similar mixing
occurs  for the fermionic partners.

\item[iii)] Couplings that deviate from the MSSM ones by  corrections
suppressed only by powers of  $\mu_0/\Lambda$ (or $m_{3/2}/\Lambda$
after SUSY breaking). These include, in particular, new quartic
couplings for the Higgses.

\item[iv)] Higher order operators ({\it e.g.} with two derivatives),
suppressed only by inverse powers of $\Lambda$. Incidentally, the
ultraviolet cut-off $\sim\Lambda$  is much smaller than in more
conventional SUSY scenarios.

\end{description}
The main effects arise in the Higgs effective potential, \ the kinetic
terms and Yukawa couplings.  We consider each in turn:

\subsection{Supersymmetric Higgs potential}

The scalar potential of a SUSY theory with a general K\"ahler
potential, $K_{\rm eff}$,  and a general gauge kinetic-function,
$f_A$, can be obtained from the general Lagrangian presented in
Appendix~A, after using $f_{AB}(\phi)=\delta_{AB}f_A(\phi)$ and
eliminating the $F$ and $D$-terms. It is given by (sum over
repeated indices is always implied)
\be  V_{\rm SUSY}={\cal{G}}^{-1}_{I J} \left({\partial W_{\rm eff}\over
\partial\phi_I}\right)^* {\partial W_{\rm eff}\over \partial\phi_J}+
{1\over 4}\left[f_{A}^{-1}(\phi)\left( {\partial K_{\rm
eff}\over\partial\phi_I} \left[{\bf t_A}\right]_{IJ}\phi_J\right)^2+
{\mathrm h.c.}\right]\ ,
\label{Vgen}
\ee
where $\phi_I$ runs over all the chiral fields of the theory
$\{\varphi_i,t,S\}$; ${\bf t_A}$ are the matrix generators of the
gauge algebra in the $\phi_J$  representation ($S$ and $t$, being
singlets, do not contribute to the $D$-terms), and
${\cal{G}}^{-1}_{IJ}$ is the inverse matrix of the K\"ahler metric
\be {\cal{G}}_{IJ}\equiv  {\partial^2 K_{\rm
eff}\over\partial\phi_I\partial\phi_J^*}\ .
\label{Gij}
\ee
With our assumptions on the $S$-dependence of $K$ (see previous
section),  ${\cal{G}}_{IJ}$ decomposes in block form, with
${\cal{G}}_{St}={\cal{G}}_{S\varphi_i}=0$. [In fact, $S$ does not
contribute as a dynamical field to the potential (\ref{Vgen}), which has been 
obtained in 
the $M_p\rightarrow\infty$ limit.] The non-trivial part of the
inverse matrix ${\cal{G}}_{IJ}^{-1}$ corresponds to the
$\{\varphi_i,t\}$ block.  In terms of $\Sigma_{\rm eff}$, this inverse
(sub)matrix is
\be {\cal{G}}^{-1}_{IJ}= {\Lambda^2\over K_{\rm eff}}\left[
\begin{array}{cc}
{\displaystyle \left[{\partial^2\Sigma_{\rm
eff}\over\partial\varphi_i\partial\varphi_j^*}\right]^{-1}} &
{\displaystyle{1\over\Lambda} \left[{\partial^2\Sigma_{\rm
eff}\over\partial\varphi_i\partial\varphi_k^*}\right]^{-1} {\partial
\Sigma_{\rm eff}\over\partial\varphi_k}}\vspace{0.2cm}\\
{\displaystyle {1\over\Lambda} {\partial \Sigma_{\rm
eff}\over\partial\varphi_k^*} \left[{\partial^2\Sigma_{\rm
eff}\over\partial\varphi_k\partial\varphi_j^*}\right]^{-1}} &
{\displaystyle 1+{1\over\Lambda^2}{\partial \Sigma_{\rm
eff}\over\partial\varphi_k^*} \left[{\partial^2\Sigma_{\rm
eff}\over\partial\varphi_k\partial\varphi_l^*}\right]^{-1} {\partial
\Sigma_{\rm eff}\over\partial\varphi_l}}
\end{array}\right]\ .
\label{Inverse}
\ee
where  $[\partial^2\Sigma_{\rm
eff}/\partial\varphi_i\partial\varphi_j^*]^{-1}$ represents the
inverse matrix of  $\partial^2\Sigma_{\rm
eff}/\partial\varphi_i\partial\varphi_j^*$.  For $\Sigma_{\rm eff}$ as
defined in (\ref{Seff}) one simply has  $[\partial^2\Sigma_{\rm
eff}/\partial\varphi_i\partial\varphi_j^*]^{-1}= \delta_{ij}$ but, for
later use, we find it convenient to leave the general expressions in
terms of derivatives of $\Sigma_{\rm eff}$. With the help of (\ref{Inverse}),
the potential (\ref{Vgen}) can be rewritten as
\bea 
V_{\rm SUSY}&=&{\Lambda^2\over K_{\rm eff}} \left\{ \left( {\partial W_{\rm
eff}\over \partial\phi_i} -{3W_{\rm eff}\over\Lambda^2} {\partial
\Sigma_{\rm eff}\over\partial\varphi_i} \right)^* \left[
{\partial^2\Sigma_{\rm
eff}\over\partial\varphi_i\partial\varphi_j^*}\right]^{-1}
\left({\partial W_{\rm eff}\over \partial\phi_j}-{3W_{\rm
eff}\over\Lambda^2} {\partial \Sigma_{\rm
eff}\over\partial\varphi_j}\right)\right.\nn\\ 
&&+  {9\over
\Lambda^2}|W_{\rm eff}|^2\left\}+ {K_{\rm eff}^2\over
4\Lambda^4}\left\{f_{A}^{-1}(\phi)\left[ {\partial \Sigma_{\rm
eff}\over\partial\varphi_i}({\bf t_A}\varphi)_i\right]^2+ {\mathrm
h.c.}\right\}\right.\ .
\label{Vgen2}
\eea
Concerning the $D$-term part of the potential, note that, if
$\Sigma_{\rm eff}$ contains terms of the form
$a(\varphi)+a(\varphi^*)$ [like the $\lambda$-dependent part of
(\ref{Seff})], gauge invariance implies 
\be 
{\partial
a(\varphi)\over\partial\varphi_i}({\bf t_A}\varphi)_i=0\ , 
\ee 
so that
$a(\varphi)$ contributes to the $D$-terms in $V$ only through the
pre-factor $K_{\rm eff}$, giving therefore only non-renormalizable
($\Lambda$-suppressed)  contributions.

The supersymmetric part of the potential for the Higgs fields,
relevant to the study of electroweak symmetry breaking, is given
[after plugging (\ref{Keff},\ref{Weff}) and $f_{A}=g_A^{-2}$  in
(\ref{Vgen2})] by
\bea V_{\rm SUSY}(H_i,t)&=&\exp\left\{-2{t+ t^* \over \Lambda}  -
{1\over\Lambda^2}\left[|H_1|^2 + |H_2|^2 +\left(\lambda H_1\cdot H_2
+{\mathrm h.c.}\right)\right]\right\}\nn\\ &\times & \left\{
-3{|\mu_0|^2\over\Lambda^2}|H_1 \cdot H_2|^2\left[
1-{6\over\Lambda^2}\left(\lambda H_1\cdot H_2  +{\mathrm
h.c.}\right)\right] \right.\nn\\ &+&\left.
|\mu_0|^2\left[\left|1-{3\over\Lambda^2}\lambda H_1\cdot H_2\right|^2
+{9\over\Lambda^4}|H_1 \cdot H_2|^2\right]  (|H_1|^2 +
|H_2|^2)\right\} \nn\\ &+& \exp\left\{-2{t+ t^* \over \Lambda}  +
{2\over\Lambda^2}\left[|H_1|^2 + |H_2|^2 +\left(\lambda H_1\cdot H_2
+{\mathrm h.c.}\right)\right]\right\}\nonumber\\ &\times &
\left\{{1\over8}(g^2+{g'}^2)\left(|H_1|^2 -|H_2|^2\right)^2
+{1\over2}g^2\left(|H_1|^2  |H_2|^2-|H_1\cdot H_2|^2
\right)^2\right\}\ , \nonumber\\ &&
\label{Veff}
\eea
where $g$ and $g'$ are the $SU(2)_L$ and $U(1)_Y$ gauge coupling
constants and
\be H_1=\left[\begin{array}{c} H_1^0\\ H_1^-
\end{array}\right]\ ,
\;\;\;\;\;\; H_2=\left[\begin{array}{c} H_2^+\\ H_2^0
\end{array}\right]\ ,
\ee
are the usual two Higgs doublets.

One noteworthy feature of this potential is that ${\rm Re}[t]$
exhibits a runaway behaviour (if $H_{1,2}$ take non-zero VEVs).  At
this level, that is inconsistent with our replacement $T=\langle
T\rangle +t/\Lambda$, which assumes $T$ is stabilized  at $\langle
T\rangle$ as given by (\ref{Tvev}). In fact, that stabilization occurs
only after SUSY is broken \cite{Pok}, as we discuss in the next
section.  For the time being, we simply ignore the problem and examine
other peculiarities of $V_{\rm SUSY}$, limiting our analysis to its
$H_i$-dependent part and ignoring the radion field.  As expected, for
$\Lambda\rightarrow \infty$ the radion field gets  decoupled and
eq.~(\ref{Veff}) reduces to the ordinary MSSM potential.  An expansion
of (\ref{Veff}) in powers of  $H_i/\Lambda$, keeping only
renormalizable terms, gives: 
\bea V(H_i)&= &|\mu_0|^2\left(|H_1|^2 +
|H_2|^2\right) +{1\over8}(g^2+{g'}^2)\left(|H_1|^2
-|H_2|^2\right)^2\nn\\ &+&{1\over2}g^2\left(|H_1|^2  |H_2|^2-|H_1\cdot
H_2|^2 \right)-{|\mu_0|^2\over \Lambda^2}\left[ \left(|H_1|^2 +
|H_2|^2\right)^2\right.\nn\\ &+&\left.  4\left(|H_1|^2 +
|H_2|^2\right)\left(\lambda H_1\cdot H_2 +{\mathrm
h.c.}\right)+3|H_1\cdot H_2|^2\frac{}{}\right]+...
\label{Vexp}
\eea Besides the usual MSSM quartic couplings, related to gauge
couplings, this potential contains additional contributions which are
purely supersymmetric and arise from the non-minimal  nature of the
K\"ahler potential and the coupling $t H_1\cdot H_2$ in $W_{\rm eff}$.

At first sight one might worry about the stability of the vacuum  with
a potential like (\ref{Vexp}). In fact, along the $D$-flat  direction
$\varphi\equiv {\rm Re}(H_1^0+H_2^0)$, the potential has a quartic
coupling that can be negative for some choices of $\lambda$: \be
V(\varphi)=\frac{1}{2}|\mu_0|^2\varphi^2-{1\over 16}
{|\mu_0|^2\over\Lambda^2} (7+8\lambda+8\lambda^*)\varphi^4+...  \ee
However, this behaviour is an artifact of the
$\varphi/\Lambda$-expansion.  The full potential (\ref{Veff}) along
the $D$-flat direction, \be  V(\varphi)= {1\over 8}|\mu_0|^2\varphi^2
\left[{9\over 2}{\varphi^2\over\Lambda^2}+
\left|2-3(1+\lambda){\varphi^2\over 2\Lambda^2}
\right|^2\right]\exp\left\{-\left[1+{1\over2}(\lambda+\lambda^*)
\right]{\varphi^2\over 2\Lambda^2}\right\} \ , \ee is not only bounded
from below but also positive definite for any value of $\lambda$, as
it should. In fact this holds for any other field direction and means,
in particular, that the electroweak symmetry is unbroken until
supersymmetry-breaking effects are considered, just like in the MSSM.

As we will discuss in section~4, the presence of new contributions to
Higgs quartic couplings has a dramatic impact on the value of the
lightest Higgs  boson mass. It would be premature at this stage to
extract implications on this respect from (\ref{Veff}) because one
expects that supersymmetry breaking physics also gives rise to new
Higgs quartic couplings of order $m_{3/2}^2/\Lambda^2$ which can
compete with the supersymmetric ones just presented.   In fact, the
presence of such adimensional  SUSY-breaking terms and its effect on
the Higgs mass have been discussed in ref.~\cite{nir} for a variety of
models with low-energy supersymmetry breaking (see also \cite{qusoft}).  
Let us remark again
that a novel feature of  our scenario is that it gives new
contributions to Higgs quartic couplings which, unlike those studied
in \cite{nir}, are truly  supersymmetric and arise from the
non-minimal K\"ahler potential.

Before closing this subsection, we mention another possible source of
non-standard supersymmetric contributions to Higgs quartic couplings,
although we do not consider it for the discussion of the
phenomenological  implications. Suppose that the original
superpotential (\ref{Wvis}) contains non-renormalizable terms like
\be 
\delta W_{\rm vis}=  {\hat c\over
\sqrt{3}M_{p}}(\hat{H}_1\cdot\hat{H}_2)^2\ .
\label{dW0}
\ee
After changing to canonically normalized fields and shifting the
radion field $T$,  we get a contribution to the effective
superpotential of the form
\be \delta W_{\rm eff}= {c \over\Lambda}(H_1\cdot
H_2)^2e^{-3t/\Lambda}\ ,
\label{dW}
\ee
with $c\equiv \hat c e^{\langle K/2M_p^2\rangle}e^{-3i\langle {\rm Im}
T\rangle}$. It is straightforward to  show that (\ref{dW}) contributes
to the effective potential (\ref{Vexp}) the additional renormalizable term
\be  \delta V_{\rm SUSY}=2\left(c{\mu_0^*\over\Lambda}H_1\cdot H_2+{\mathrm
h.c.}\right) \left(|H_1|^2 + |H_2|^2\right)\ ,  \ee
which is of the same order as that of the quartic terms considered in
(\ref{Vexp}), unless $c$ is very small.

\subsection{Kinetic terms, contributions to the $\rho$-parameter}

A non-minimal K\"ahler potential like (\ref{Keff}) leads, after
electroweak symmetry breaking, to non-canonical kinetic terms for
scalar fields (see Appendix~A):
\be
{\cal{L}}_{\rm kin}= \langle {\cal{G}}_{I J}\rangle  D_\mu \phi_I
D^\mu \phi^*_J\ ,
\label{LK}
\ee 
where the brackets indicate that all fields are replaced by their
VEVs.  $\langle{\cal{G}}_{I J}\rangle$ is in general a  non-diagonal
matrix, in particular for the entries involving  the Higgses and the
radion, due to ${\cal{O}}(v/\Lambda)$ terms:
\be \langle{\cal{G}}_{IJ}\rangle=\delta_{IJ}+{\cal{O}}
\left({v\over\Lambda}\right)\ , \ee
where $v$ denote the Higgs VEVs.  Getting the kinetic terms back to
canonical  form requires a re-scaling and redefinition of fields,
which affects  other couplings in the Lagrangian. The implications of
such effects on the composition and masses of the Higgs fields are
discussed in detail in section~6 and Appendix~C.  There are also
implications for gauge boson masses.  Explicitly, (\ref{LK}) gives,
for the Higgs kinetic terms: \bea {\cal{L}}_{\rm kin}&=&{\langle
K_{\rm eff}\rangle\over\Lambda^2}\left[ |D_\mu H_1|^2+|D_\mu
H_2|^2\right]\nn\\ &&+{\langle K_{\rm eff}\rangle\over\Lambda^4}\left|
(\overline{H}_1+\lambda H_2)\cdot D_\mu H_1+ (\overline{H}_2-\lambda
H_1)\cdot D_\mu H_2 \right|^2\ ,
\label{LKp}
\eea where $K_{\rm eff}$ is given by eq.~(\ref{Keff}) and
$\overline{H}_i$ is the $SU(2)$-conjugate of $H_i$
($\overline{H}_i\equiv -i{\bma \sigma}_2H_i^*$;  with $|H_i|^2\equiv
H_i\cdot\overline{H}_i$).  From the expression above, it is
straightforward to obtain the gauge boson masses: \be M_Z^2={1\over
4}(g^2+{g'}^2)v^2{\langle K_{\rm eff}\rangle\over\Lambda^2}
\left[1+{v^2\over2\Lambda^2}\cos^22\beta\right]\ ,
\label{MZ}
\ee  and
\be M_W^2={1\over 4}g^2v^2{\langle K_{\rm eff}\rangle\over\Lambda^2}\ ,
\label{MW}
\ee
with $v^2=2(|H_1|^2+|H_2|^2)$, $\tan\beta\equiv\langle
H_2^0\rangle/\langle H_1^0\rangle$ and
\be  {\langle K_{\rm
eff}\rangle\over\Lambda^2}=1+2(1+\lambda\sin2\beta){v^2\over\Lambda^2}+
{\cal{O}}\left({v^4\over\Lambda^4}\right)\ .   \ee
{}From the masses (\ref{MZ}) and (\ref{MW}) we first see that there is
a deviation of $v$  from its minimal value $\sim 246$ GeV. That
deviation is a small ${\cal{O}}(v^2/\Lambda^2)$  effect that we
neglect in the following. Then we see that there is also a tree-level
deviation from $\rho=1$, with $\Delta\rho\simeq
-\frac{1}{2}(v^2/\Lambda^2)\cos^22\beta$, which can be traced back to
the  terms in the second line of (\ref{LKp}).  Consequently, the
comparison to the present measurements of the $\rho$ parameter
translates into a bound on $\Lambda$. The precise value of the latter
depends on the value of the Higgs mass, $M_{h^0}$, because the larger
$M_{h^0}$, the lower the SM prediction for $\rho$. Since the above tree-level
contribution is negative, the corresponding bound on $\Lambda$ becomes
even stronger for larger $M_{h^0}$. In order to give a more quantitative
estimate of the bound, notice that the $2\sigma$
experimental window (with $M_h^0$ unconstrained), 
$\Delta\rho=0.9998^{+0.0034}_{-0.0012}$ \cite{PDG},  
leaves little room for non SM contributions, and we get the bound
\be
\label{ro-bound}
\Lambda \gsim (5.5\ {\rm TeV}) |\cos 2\beta|\ .
\ee
As we will show in section~4, a correct breaking of the electroweak symmetry
in this context normally demands $\tan\beta\simeq 1$ and, therefore, 
this source of non-zero $\Delta\rho$ and the previous upper bound disappear.

The tree-level behaviour of $\rho$ for $\tan\beta=1$ can be understood 
by symmetry arguments. Let us define the bi-doublets
\be
{\bf H}=\left[
\begin{array}{cc}
H_1^0 & H_2^+ \\
H_1^- & H_2^0
\end{array}\right]\ ,\;\;\;\;
\overline{\bf H}=\left[
\begin{array}{cc}
H_2^{0*} & -H_1^+ \\
-H_2^- & H_1^{0*}
\end{array}\right]\ ,
\label{bidoublet}
\ee
built out of the two Higgs doublets $H_1$ and $H_2$ (and their conjugates
$\overline{H}_{1,2}$). In terms of ${\bf H}$ 
and $\overline{\bf H}$, the kinetic Lagrangian (\ref{LKp}) can be written as
\be
{\cal{L}}_{\rm kin}={\langle K_{\rm eff}\rangle\over\Lambda^2}{\mathrm Tr}\left[
\left(D_\mu {\bf H}\right)^\dagger
D^\mu {\bf H}\right]+
{\langle K_{\rm eff}\rangle\over\Lambda^4}
\left|{\mathrm Tr}\left[\left({\bf H}-
\lambda\overline{\bf H}\right)^\dagger D_\mu {\bf H}\right]\right|^2\ .
\ee
This expression makes manifest the approximate invariance of ${\cal{L}}_{\rm kin}$
under the $SU(2)_L\times SU(2)_R$ (global) group of transformations:
\bea
{\bf H}&\rightarrow & {\bf U_L}{\bf H}{\bf U_R}^\dagger\ ,
\label{sulr}\\
{\bf W}_\mu\equiv  {\bma\sigma}^a W^a_\mu &\rightarrow & 
 {\bf U_L}{\bf W}_\mu{\bf U_L}^\dagger\ ,
\label{sulrW}
\eea
where ${\bf U_{L,R}}$ are generic $2\times 2$ $SU(2)$ matrices
(and ${\bma\sigma}^a$ are the Pauli matrices).
To see this, notice that (\ref{sulr}) implies that
$\overline{\bf H}$ transforms as ${\bf H}$:
\be
\overline{\bf H}\rightarrow  {\bf U_L}\overline{\bf H}{\bf U_R}^\dagger\ ,
\ee
while (\ref{sulrW}) implies that, for $g'=0$,
\be 
D_\mu{\bf H}\rightarrow  {\bf U_L}(D_\mu{\bf H}){\bf U_R}^\dagger\ .
\label{sulrd}
\ee
This approximate $SU(2)_L\times SU(2)_R$ symmetry
is spontaneously broken by the VEVs of $H_1^0$
and $H_2^0$ down to a subgroup $G$. 
In the case $\langle H_1^0\rangle=\langle H_2^0\rangle$,
{\it i.e.} for $\tan\beta=1$, the unbroken symmetry $G$ is the diagonal 
$SU(2)_{L+R}$ (with ${\bf U_L}={\bf U_R}$), the so-called custodial 
symmetry \cite{custodial}.
[The generator of the unbroken electromagnetic $U(1)_Q$
corresponds to $\sigma_3$ of this $SU(2)$.]
The electroweak-symmetry-breaking 
Goldstone bosons, $\{G^+,G^0,G^-\}$, transform as a triplet under this custodial 
symmetry, ensuring $\rho=1$. This symmetry plays an important role for
the Higgs spectrum as we discuss in detail in section~6.
If $\langle H_1^0\rangle\neq\langle H_2^0\rangle$,
the unbroken subgroup $G$ is simply $U(1)_Q$ and $\rho\neq 1$.

\subsection{Yukawa couplings, fermion masses}

At the renormalizable level, the only difference in fermionic
couplings in our Lagrangian with respect to the MSSM one is the
presence of Yukawa couplings that mix the standard Higgsinos with the
fermionic component of the radion superfield (the ``radino''). From the
general Lagrangian written in Appendix~A, after eliminating the
auxiliary fields, we find, for the two-fermion couplings (with no
derivatives):
\be \delta{\cal{L}}_{\rm SUSY}={1\over 2}\left\{ -{\partial^2W_{\rm
eff}\over\partial\phi_I\partial\phi_J}+ {\partial^3K_{\rm
eff}\over\partial\phi_I\partial\phi_J\partial\phi_L^*}
{\cal{G}}^{-1}_{LK}{\partial W_{\rm eff}\over\partial\phi_K}
\right\}(\chi_I\cdot\chi_J)+{\mathrm h.c.}\ ,
\label{fermion}
\ee
where the Weyl spinors $\{\chi_I\}=\{\chi_i,\chi_t\}$ are the
fermionic partners of the scalars 
$\{\phi_I\}=\{\varphi_i,t\}$, and
$(\chi_I\cdot\chi_J)=i\chi_I^T{\bma\sigma}_2\chi_J$.  Using
(\ref{Inverse}) we can rewrite the previous expression in the form
\bea 
\delta{\cal{L}}_{\rm SUSY}&=&{1\over 2}\left\{ -{\partial^2W_{\rm
eff}\over\partial\varphi_i\partial\varphi_j}+ {2\over\Lambda}{\partial
W_{\rm eff}\over\partial\varphi_i} {\partial \Sigma_{\rm
eff}\over\partial\varphi_j} + {3W_{\rm eff}\over\Lambda^2}\left[
{\partial^2 \Sigma_{\rm eff}\over\partial\varphi_i\partial\varphi_j}
-{1\over\Lambda^2}{\partial \Sigma_{\rm eff}\over\partial\varphi_i}
{\partial \Sigma_{\rm eff}\over\partial\varphi_j}\right] \right.\nn\\
&+&\left.{\partial^3\Sigma_{\rm
eff}\over\partial\varphi_i\partial\varphi_j\partial\varphi_l^*}
\left[{\partial^2\Sigma_{\rm
eff}\over\partial\varphi_l\partial\varphi_k^*}\right]^{-1}
\left({\partial W_{\rm eff}\over\partial\varphi_k} -{3W_{\rm
eff}\over\Lambda^2}{\partial \Sigma_{\rm eff}\over\partial\varphi_k}
\right)\right\}(\chi_i\cdot\chi_j)\nn\\ &+&{2\over\Lambda}{\partial
W_{\rm eff}\over\partial\varphi_i}(\chi_i\cdot\chi_t) -{3W_{\rm
eff}\over\Lambda^2}(\chi_t\cdot\chi_t)+{\mathrm h.c.}
\label{Lfer}
\eea
Explicitly, this leads to the following Yukawa couplings (and
fermionic mass terms): \bea
\delta{\cal{L}}_{\rm 
SUSY}&=&\mu_0e^{-3t/\Lambda}\left[\chi_{H_1^-}\cdot\chi_{H_2^+}-
\chi_{H_1^0}\cdot\chi_{H_2^0}\right.\nn\\
&+&\left.{2\over\Lambda}\left( H_2^0\chi_{H_1^0}+H_1^0\chi_{H_2^0}-
H_2^+\chi_{H_1^-}-H_1^-\chi_{H_2^+} \right)\cdot \chi_t
\right]+{\mathrm h.c.}
\label{Yuk}
\eea in addition to the MSSM $h_{U,D}$-Yukawa couplings which we do
not write.  After electroweak symmetry breaking, (\ref{Yuk}) implies
that the $5\times 5$  neutralino mass matrix contains mixing terms
between $\chi_{H_i^0}$ and $\chi_t$.  To write the complete matrix we
need to introduce also the effects from SUSY breaking, which are most
important for gauginos and $\chi_t$. These effects are discussed in
the next section.

\section{Supersymmetry breaking}
\setcounter{equation}{0}
\renewcommand{\theequation}{4.\arabic{equation}}

\subsection{The mechanism of SUSY breaking}

As mentioned in section 2, supersymmetry breaking plays an essential
role in the stabilization of the radion field. In addition, it is
mandatory for a correct low-energy phenomenology. In the framework of
ref.~\cite{Pok},  the breaking of SUSY is triggered by a brane
superpotential of the form
\be W_{sbr}=W_h+e^{-3 T}W_v\ ,
\label{Wsource}
\ee
where $W_h$ and $W_v$ are constant brane sources (at $x^5=0$ and
$x^5=\pi r_0$, respectively). As was shown there, the radion field
gets stabilized in this process, although with a non-vanishing
(negative) cosmological constant in the 4-D effective theory. 
On the other hand, the dilaton field
presents a runaway behaviour. These are shortcomings for a viable
phenomenology.

Here we will discuss the above-mentioned issues: SUSY breaking,
stabilization of  $T$  and $S$, and the value of the cosmological
constant. In the next subsection we will extract  and discuss the form
of the soft terms.  We will not make any initial assumptions about the
possible dependence of $W_{sbr}$ on $S$ and the form of
$K(S,S^\dagger)$. Actually, $W_{sbr}$ will have a non-trivial
$S$-dependence if it is generated by  gaugino condensation (recall
that the ordinary dilaton is intimately related to the gauge coupling
constant). On the other hand, in the context of string theory we
expect large perturbative and  non-perturbative corrections to the
K\"ahler potential \cite{kahlercorr}.

To study gravity-mediated SUSY-breaking effects we have to  consider
the whole SUGRA potential [rather than (\ref{Vgen})].  First, to
discuss how $W_{sbr}$ breaks SUSY we consider the  SUGRA effective
potential for $T$ and $S$, neglecting matter fields  (the effects of
electroweak symmetry breaking  represent small  corrections in this
analysis).  This potential reads \cite{Cremmer}
\be V_{\rm SUGRA}(T,S)=\exp(K/M_p^2)\left[\sum_{I,J=T,S} L_I^*
{\cal{G}}_{IJ}^{-1}L_J-3{|W|^2\over M_p^2} \right]\ ,
\label{Vsugra}
\ee
with $K$ the K\"ahler potential, ${\cal{G}}_{IJ}$ its associated
metric [as defined by (\ref{Gij})],  $W$ the superpotential
(\ref{Wsource}),  and
\be L_J\equiv {\partial W\over\partial\phi_J}+{W\over M_p^2}{\partial
K \over \partial\phi_J}\ .
\label{Lm}
\ee
Alternatively,  \be V_{\rm SUGRA}(T,S)=\sum_{I,J=T,S} F_I {\cal{G}}_{IJ}
F_J^*-3 {|W|^2\over M_p^2}\exp(K/M_p^2)\ ,
\label{Vsugra2}
\ee
with  
\be 
F_I\equiv \exp(K/2M_p^2) L_K^*  {\cal{G}}_{KI}^{-1}\ .  
\ee
Provided $K$ can be decomposed as $K=K(S,S^\dagger)+K(T,T^\dagger)$
(which is the most plausible case, as argued in sect.~2) and $W_{sbr}$
has a factorizable dependence on $S$, $V$ presents a stationary point
in $T$ at
\be
\label{FT}
F_T\equiv  \exp[K/(2M_p^2)]{\cal{G}}_{TT}^{-1}L_T^*=0\ .  \ee
This stationary point will often correspond to a minimum in $T$. This
is guaranteed, in particular, if $V$  is vanishing at that
point. Using the explicit form for  $K(T,T^\dagger)$ given in
Eq.~(\ref{EffK2}),  then Eq.~(\ref{FT}) is equivalent to the
minimization condition $\partial V/\partial T=0$, which gives
\cite{Pok}:
\be W_v\exp[-3\langle T\rangle]+W_h\exp[-\langle T  +
T^\dagger\rangle]=0\ .
\label{minT}
\ee
Now, the parameters in the brane superpotential (\ref{Wsource})  can
be arranged so as to obtain  the necessary hierarchy (\ref{Tvev}),
that is, $\exp[-\langle T + T^\dagger\rangle]=\Lambda^2/(3M_p^2)$,
which follows if $\sqrt{3}|W_h| M_p=|W_v|\Lambda$.  In addition, one
should have  $W_v\sim M_p^3$, so that $\langle W_{sbr}\rangle$ has the
right order of magnitude for  a sensible gravitino mass, as we shall
see shortly.  The VEV for the imaginary part of $T$ is determined
by the relative phase between $W_h$ and~$W_v$.

The stabilization of $S$ is not so straightforward: as mentioned
above, for $W_h$, $W_v$ constant and
$K(S,S^\dagger)=-M_p^2\log(S+S^\dagger)$, $S$ exhibits a runaway
behaviour. If, following ref.~\cite{Pok}, one fixes the dilaton by
hand at a given value, then the potential has the minimum at negative
$V$, which jeopardizes a sensible phenomenology. So, in order to
proceed we have to  modify the scenario so that $S$ may be stabilized
at a finite value   by some unknown mechanism while the cosmological
constant vanishes. This could be accomplished in several ways.  For
instance, if $K=K(S,S^\dagger)+K(T,T^\dagger)$  with
$K(S,S^\dagger)=-3M_p^2\ln(S+S^\dagger)$,  as in no-scale models
\cite{noscale}, this choice guarantees a zero cosmological constant
for all values of $S$ [which means in particular that the  condition
$F_T=0$, eq.~(\ref{FT}), corresponds to a true minimum].  This form of
$K(S,S^\dagger)$ is perfectly plausible. Recall that $S$ may represent
the ordinary dilaton or any other moduli field living in the bulk. In
particular, in the context of string theory, it may represent one of
the moduli which parametrize the size and shape of the six extra
compactified dimensions. In the overall modulus approximation
(i.e. all extra radii equal) the  K\"ahler potential is exactly as the
one required \cite{Witten:1985xb}.  There is however a problem related to this
scenario. Namely,  the potential has flat directions along the real
and imaginary parts of $S$,  with the corresponding massless particles
in the spectrum. Eventually, non-perturbative corrections could lift
these flat directions, fix $\langle S\rangle$ and give a (light) mass
to these fields. As will be shown shortly,  the couplings of the $S$
field to matter fields are suppressed by  inverse powers of $M_p$. The
presence of such light weakly-interacting scalar fields could be very
dangerous for a viable cosmology \cite{cosmotrouble}, although,
definitely, the subject deserves  further study.

Here we take a different route and assume that (\ref{Wsource}) has
also an $S$-dependence (at this point we deviate from the assumptions
of \cite{Pok}; for alternative options see \cite{Lalak}) while keeping 
$K(S,S^\dagger)= -M_p^2\ln(S+S^\dagger)$.\footnote{Although we do not 
continue analyzing  here the no-scale
possibility, we will see at the end of the section that it  leads to
results very similar to those arising from the present assumptions.}
Since $F_T=0$, the condition for $V(\langle T\rangle, \langle
S\rangle)=0$ reads
\be
\label{Vcero}
\left\langle\left|(S+S^\dagger){\partial W\over\partial S }-W\right|^2
\right\rangle =3\left\langle |W|^2\right\rangle\ .  \ee
If the dependence of $W_{sbr}$ on $S$ is factorizable\footnote{The
hypothesis of factorizability is by no means necessary, but it is
convenient since, as discussed above, it guarantees $F_T=0$ at the
minimum, preserving the cancellation of the cosmological constant. On
the other hand, notice that in this case the  $S$-dependence of $W$
may be moved into $K$ by a K\"ahler  transformation.},
i.e. $W_{sbr}\propto f(S)$, then eq.~(\ref{Vcero}) is equivalent to
\be \left\langle (S+S^\dagger){1\over f}{\partial f\over\partial S}
\right\rangle\equiv\varsigma=1+\sqrt{3}e^{i\alpha}\ ,
\label{varsigma}
\ee
with $\alpha$ an arbitrary phase.  We will assume this in the
following.  Obviously we are not solving the cosmological constant
problem  but simply making the necessary adjustments to ensure such
cancellation.

In summary, we are therefore led to a scenario with supersymmetry breaking driven
by the dilaton, with \be F_S\equiv
\exp[K/(2M_p^2)]{\cal{G}}_{SS}^{-1}L_S^*\neq 0\ , \ee and cosmological
constant tuned to zero. The resulting gravitino mass is \be m_{3/2}=
\exp[\langle K\rangle/(2M_p^2)] {|\langle W\rangle |\over M_p^2}\sim
{| W_h|\over M_p^2}\sim \Lambda\ , \ee of the correct order of
magnitude.

\subsection{Soft breaking terms}

Supersymmetry breaking is communicated to the matter fields in the
visible brane by gravitational mediation. The resulting soft masses
will be given in terms of the two (complex) quantities: $W_h$ and
${\widetilde{W}}_v\equiv W_v \exp[-3\langle T\rangle]$.  These two
sources of SUSY breaking are related by the minimization condition
(\ref{minT}), so that if we write $W_h=|W_h|e^{i\gamma}$,  then
${\widetilde{W}}_v=-|{\widetilde{W}}_v|e^{i\gamma}$.  All soft masses
can then be given in terms of the complex mass
\be  {\tilde{m}}_{3/2}\equiv m_{3/2}e^{i\gamma}\ .  \ee
Although $m_{3/2}$ is dominated by the first source of SUSY breaking
($W_h$),  in general both sources can give comparable contributions to
other soft masses.

The effects of soft SUSY breaking in masses and other couplings of
scalar observable fields, $\varphi_i$, are obtained from the SUGRA
effective potential (\ref{Vsugra}) where now we include also matter
fields in the sum, i.e. $I,J= \varphi_i,T,S$, and use the complete
effective  superpotential
\bea W &=& e^{\langle K/2M_p^2\rangle}W_{sbr}\ +\ W_{\rm eff}
\nonumber\\ &=&
\tilde{m}_{3/2}\left[M_p^2-{1\over3}\Lambda^2e^{-3t/\Lambda}\right]
{f(S)\over f(\langle S\rangle)}\  +\ W_{\rm eff}\ , \eea
where  $W_{\rm eff}$ is the effective superpotential for the matter
fields,  given in (\ref{Weff}), and we have made the shift $T=\langle
T\rangle +t/\Lambda$  to the physical field $t$. The corresponding
shift for $S$ is  $S=\langle S\rangle + 2s\langle{\rm
Re}S\rangle/M_p$. It turns out that the interactions of the shifted
field $s$ to matter fields are always suppressed by inverse powers of
$M_p$. This is, essentially, a consequence of its normalization
factor, $\sim M_p^{-1}$ (note that the normalization factor for $t$ is $\sim
\Lambda^{-1}$). Hence, we concentrate here in the soft-terms  for
$\{\phi_I\}=\{\varphi_i,t\}$ only.  We can conveniently write these
contributions in terms of $K_{\rm eff}$, $\Sigma_{\rm eff}$  and
$W_{\rm eff}$ as defined in eqs.~(\ref{Keff}--\ref{Weff}).  Making an
expansion in powers of $M_p^{-1}$ we find, for the $L_J$'s defined in
(\ref{Lm}):
\bea
\label{Ls}
L_S&=&(\varsigma-1){\tilde
m}_{3/2}\left[M_p-{1\over3M_p}\Lambda^2e^{-3t/\Lambda}\right]- {W_{\rm
eff}\over M_p}+...\\ L_t&=&{\tilde
m}_{3/2}\Lambda\left[e^{-3t/\Lambda}-{K_{\rm
eff}\over\Lambda^2}\right] -3{W_{\rm eff}\over \Lambda}+...\\
L_{\varphi_i}&=&{\partial W_{\rm eff}\over\partial\varphi_i}+{{\tilde
m}_{3/2}\over \Lambda^2} K_{\rm eff}{\partial\Sigma_{\rm
eff}\over\partial\varphi_i}+...
\label{Lvarphi}
\eea
Notice that, prior to electroweak breaking,  $\langle
L_S\rangle=(\varsigma-1){\tilde m}_{3/2}M_p$ and $\langle L_t\rangle=
\langle L_{\varphi_i}\rangle=0$, corresponding to the fact that it is
$F_S$ that breaks SUSY.  After substituting (\ref{Ls}--\ref{Lvarphi})
in the SUGRA potential, keeping only terms not suppressed by
$M_p^{-1}$, this can be expressed in the following suggestive form,
\be V_{\rm SUGRA}={\cal{G}}_{IJ}^{-1} \left({\partial {\cal{W}}_{\rm
eff}\over \partial\phi_I}\right)^* {\partial {\cal{W}}_{\rm eff}\over
\partial\phi_J}-\left[(2+\varsigma) {\tilde m}_{3/2}^*W_{\rm eff}+
{\mathrm h.c.}\right]\ ,
\label{softb}
\ee
%
where we have introduced the non-holomorphic function
\be {\cal{W}}_{\rm eff}\equiv W_{\rm eff}
-{1\over3}{\tilde{m}}_{3/2}\Lambda^2e^{-3t/\Lambda}
+{\tilde{m}}_{3/2}K_{\rm eff}\ .
\label{calWeff}
\ee
The first term in the right hand side of (\ref{softb}) is exactly of
the form of a SUSY potential. However, ${\cal{W}}_{\rm eff}$, being
non-holomorphic, cannot  be interpreted as a superpotential.
Nevertheless, it is instructive to identify the holomorphic pieces of
${\cal{W}}_{\rm eff}$ which may be interpreted as an effective
superpotential $W_{\rm eff}'$. In an expansion in powers of
$\Lambda^{-1}$, keeping only renormalizable terms, we get
\bea  W_{\rm eff}'&=&{2\over 3}{\tilde{m}}_{3/2}\Lambda^2+(\mu_0
+\lambda {\tilde{m}}_{3/2})H_1\cdot H_2 +h_{U}^{ij} Q_{Li}\cdot  H_2
U_{Rj} + h_{D}^{ij} H_1\cdot Q_{Li}  D_{Rj}\nn\\
&-&{\tilde{m}}_{3/2}\Lambda t^2-{1\over\Lambda}(3\mu_0 +\lambda
{\tilde{m}}_{3/2})tH_1\cdot H_2 +{4\over
3}{{\tilde{m}}_{3/2}\over\Lambda}t^3\ .
\label{Weffp}
\eea
This corresponds to a NMSSM superpotential, where the $t$ field plays
the role of the NMSSM singlet, with a non-zero $\mu$-term (while the
usual NMSSM assumes it is absent to start with and is generated
dynamically by the singlet VEV). However, the model cannot be
interpreted as a particular version of the NMSSM for several reasons: 
the K\"ahler metric ${\cal{G}}_{IJ}$ in
eq.(\ref{softb}) derives  from a non-minimal  K\"ahler potential,
$K_{\rm eff}$; and, besides  the ``supersymmetric'' part,
eq.(\ref{softb}) contains extra pieces [the last term plus the
non-holomorphic pieces in ${\cal{W}}_{\rm eff}$, which are not
accounted for by (\ref{Weffp})],  which correspond to SUSY breaking
terms. These terms are more general than those usually considered
for the NMSSM ({\it e.g.} SUSY-breaking quartic Higgs couplings are 
present).
Consequently, the expected  phenomenology will be completely different.

Let us stress that eq.(\ref{Weffp}) shows explicitly  the two sources
for the $\mu$ parameter:
\be \mu\equiv \mu_0 +\lambda \tilde{m}_{3/2}.
\label{mucomp}
\ee
One is simply the previously mentioned supersymmetric Higgs mass  of
order ${\cal O}(M_{p})$ in the observable superpotential, which is
effectively reduced to ${\cal O}(\Lambda)$ by the warp factor; the
other is the $[\lambda H_1\cdot H_2 +{\mathrm h.c.}]$ term in the
K\"ahler potential (\ref{Keff}) that, after SUSY breaking, contributes
to the $\mu$ parameter via the Giudice-Masiero \cite{GM} mechanism.

We can simplify the potential (\ref{softb}) by using the explicit form
of ${\cal{G}}_{IJ}^{-1}$ given in (\ref{Inverse}).  We find soft terms
in the scalar potential proportional to $m_{3/2}^2$ of the form
\bea \delta V_{\rm soft}&=&m_{3/2}^2\left\{{\Lambda^2\over K_{\rm
eff}}\left[ \Lambda^2+{\partial\Sigma_{\rm
eff}\over\partial\varphi_i^*} \left[{\partial^2\Sigma_{\rm
eff}\over\partial\varphi_i\partial\varphi_j^*}\right]^{-1}
{\partial\Sigma_{\rm eff}\over\partial\varphi_j} \right]
e^{-3(t+t^*)/\Lambda} \right.\nonumber\\ &-&\left.\Lambda^2\left[
e^{-3t/\Lambda}+e^{-3t^*/\Lambda}\right]+K_{\rm eff}
\frac{}{}\right\}\ .
\label{softsquared}
\eea
plus soft terms which are linear in the gravitino mass:  \bea \delta
V_{\rm soft}&=&\tilde{m}_{3/2}\left\{{\Lambda^2\over K_{\rm eff}}\left(
{\partial W_{\rm eff}\over \partial\varphi_i}-{3W_{\rm eff}\over
\Lambda^2}{\partial\Sigma_{\rm eff} \over \partial\varphi_i}\right)^*
\left[{\partial^2\Sigma_{\rm
eff}\over\partial\varphi_i\partial\varphi_j^*}\right]^{-1}
{\partial\Sigma_{\rm eff}\over\partial\varphi_j}
e^{-3t/\Lambda}\right.\nonumber\\ &+&\left.W_{\rm eff}^*
\left[1-\varsigma-{3 \Lambda^2\over K_{\rm eff}}e^{-3t/\Lambda}\right]
\frac{}{}\right\}+{\mathrm h.c.}
\label{softlinear}
\eea An expansion of (\ref{softsquared}) and (\ref{softlinear}) in
powers of $\varphi_i/\Lambda$ and $t/\Lambda$ makes manifest the
properties of the SUSY breaking terms in this scenario.  In increasing
order of field powers, we find:
\begin{description}

\item[a)] vanishing field-independent part of $V_{\rm soft}$.

\item[b)] vanishing $t$-tadpole.

\item[c)] universal soft mass (equal to $m_{3/2}$) for the scalar
fields $\varphi_i$'s.

\item[d)] the real part of the radion field has a soft mass $m_{3/2}$
while its imaginary part has mass $3m_{3/2}$.

\item[e)] $B$-type and $A$-type soft terms. More precisely, the
$A$-parameters are universal, $A^*=\tilde{m}_{3/2}(1-\varsigma)$,
while  $B\mu=(-\varsigma^*\tilde{m}_{3/2}^*)\mu_0 +
(2\tilde{m}_{3/2}^*)\lambda \tilde{m}_{3/2}$, where $\varsigma$ and
$\mu$ are given by eqs.(\ref{varsigma}, \ref{mucomp}). The latter
expression reflects the two different contributions to the
$B$-parameter, associated with the two sources of $\mu$.

\item[f)] trilinear terms like $t^*|\varphi|^2\ +\ {\rm h.c.}$,  where
$\varphi$ is any chiral scalar field.  These terms always involve the
radion field, $t$.

\item[g)] dimension $4+n$ SUSY-breaking operators suppressed
only by powers of $m_{3/2}/\Lambda^{1+n}$.

\end{description}
The points a), b) above are a consequence of our previous assumption
of vanishing cosmological constant and the fact that we have defined
$t$ as the physical field that describes fluctuations of $T$ around
its tree-level minimum $\langle T\rangle$ (arising from $F_T=0$).  The
other points are more or less conventional, except for points f) and
g). Actually, one may worry  about the quadratic divergences
introduced by the dimension $D\geq 4$ operators, with undesirable
consequences  for the gauge-hierarchy. Certainly, non-supersymmetric
$D=4$ operators give a non-zero contribution to the field-independent
part of  ${\rm Str} {\cal M}^2$, and thus to the
quadratically-divergent part of the one-loop effective potential
\bea V_{\rm 1-loop}^{\rm quad} = \frac{1}{32\pi^2}\Lambda_{UV}^2  {\rm
Str} {\cal M}^2\ ,
\label{Vquad}
\eea
where $\Lambda_{UV}$ is the ultraviolet cut-off of the effective
theory. As discussed in sect.~2, $\Lambda_{UV}\simeq \Lambda ={\cal
O}({\rm TeV})$.  One can think that, provided $\Lambda$ is not much
bigger than 1 TeV,  the contribution (\ref{Vquad}) will be small
enough not to spoil the naturalness of the electroweak breaking. But,
actually, even for $\Lambda\gg 1$ TeV  the contribution to $V_{\rm
1-loop}^{\rm quad}$ remains under control. The reason is that, for a
renormalizable $W_{\rm eff}$, all the $D\geq 4$ operators that
contribute to (\ref{Vquad})  contain at least a factor $\Lambda^{-2}$
that precisely compensates the  $\Lambda_{UV}^{2}$ factor\footnote{If
$W_{\rm eff}$ contains non-renormalizable operators, like the one in
eq.~(\ref{dW}), this is no longer true. Such a term in the
superpotential induces a term $\sim (m_{3/2}/\Lambda) \phi^4$ after
SUSY breaking. However this term is still harmless, as it gives a
vanishing contribution to ${\rm Str} {\cal M}^2$ (in this sense it is
a ``soft'' term).}  (incidentally, this also occurs for the
supersymmetric $D\geq 4$ operators of sect.~2).  Then, provided
$m_{3/2}$ and $\mu$ are of order ${\cal O}({\rm TeV})$,  $V_{\rm
1-loop}^{\rm quad}$ is small, thanks to the additional suppression
provided by the 1-loop $1/32\pi^2$ factor.

Soft gaugino masses are also generated and given by
\be M_A=\frac{1}{2}F_i{\partial {\rm Re} f_{A}\over \partial \phi_i}\
, \ee
($A$ is a gauge index).  In our case, SUSY is broken along the $S$
direction, so that $M_A$ will be sizeable [$\sim {\cal{O}}(\Lambda)$]
as long as the gauge kinetic functions, $f_A$, have a non-trivial
$S$-dependence. Of course, this is so in the most ordinary case,
i.e. when $S$ represents the dilaton and $f_A \sim S$. Even if one
departs from this simple situation (and we argued in favor of such a
case in sect.~2), $f_A$ will generically present a non-trivial
$S$-dependence, thus guaranteeing sizeable gaugino masses. {\it E.g.} if
$S$
represents a modulus, $U$, parameterizing one of the three complex
extra dimensions, $f_A$ may have a non-trivial dependence on $U$. This
occurs in the context of $M$-theory compactified \`a la Horava-Witten
\cite{Horava-Witten} and for appropriate 5-branes in the context of 
Type I constructions \cite{TypeI}. 
It may also occur in the weakly-coupled heterotic string if the
threshold corrections are large. We prefer to leave $M_A$ as a free
parameter since its precise value depends on this kind of details. In
fact, in many other constructions $f_A$ has important contributions
which are model- and gauge-group-dependent, and prevent the writing of
a general expression for $f_A$ ({\it e.g.} in Type I constructions there
are
model-dependent  contributions proportional to the twisted moduli).

Finally, we discuss the effects of SUSY breaking for the masses of
matter fermions $\chi_I=\{\chi_i,\chi_t\}$. From the SUGRA Lagrangian
we obtain (after eliminating the mixing between matter fermions and
the gravitino  \cite{WessBagger}): 
\bea 
\delta
{\cal{L}}_{\rm SUGRA}&=&{1\over2}\exp[K/(2M_p^2)]{|W|\over W}\left\{
{\partial L_J\over\partial\phi_I}- {\partial^3
K\over\partial\phi_I\partial\phi_J\partial\phi_L^*}
{\cal{G}}_{LK}^{-1}L_K\right.\nn\\ &+&\left.{1\over W}\left[-{\partial
W\over\partial\phi_I}L_J +{1\over 3}L_IL_J\right]
\right\}(\chi_I\cdot\chi_J)+{\mathrm h.c.}  
\eea 
where $L_I$ was
defined in (\ref{Lm}). Substituting the $L_I$ expansions as given in
(\ref{Ls}--\ref{Lvarphi}) we arrive at the $M_p\rightarrow\infty$ limit
\be \delta{\cal{L}}_{\rm SUGRA}\rightarrow{1\over 2}\left\{
-{\partial^2{\cal{W}}_{\rm eff}\over\partial\phi_I\partial\phi_J}+
{\partial^3K_{\rm
eff}\over\partial\phi_I\partial\phi_J\partial\phi_L^*}
{\cal{G}}_{LK}^{-1}{\partial {\cal{W}}_{\rm eff}\over\partial\phi_K}
\right\}(\chi_I\cdot\chi_J)+{\mathrm h.c.}\ ,
\label{softfer}
\ee
with ${\cal{W}}_{\rm eff}$ as defined in (\ref{calWeff}). We see once
again that (\ref{softfer}) has a supersymmetric form [compare with
(\ref{fermion})] but is not supersymmetric because ${\cal{W}}_{\rm
eff}$ is a non-holomorphic function of the chiral fields.

Rewritting (\ref{softfer}) in terms of $\Sigma_{\rm eff}$ using
(\ref{Inverse}) and explicitly replacing $\phi_I=\{\varphi_i,t\}$,
$\chi_I=\{\chi_i,\chi_t\}$,  we obtain the following
${\tilde{m}}_{3/2}$-dependent terms:
\bea  \delta{\cal{L}}_{\rm soft}&=&{1\over
2}{\tilde{m}}_{3/2}e^{-3t/\Lambda}\left\{ \left[{1\over\Lambda^2}
{\partial \Sigma_{\rm eff}\over\partial\varphi_i} {\partial
\Sigma_{\rm eff}\over\partial\varphi_j} -{\partial^2 \Sigma_{\rm
eff}\over\partial\varphi_i\partial\varphi_j} \right.\right.\nn\\
&+&\left. \left.  {\partial^3\Sigma_{\rm
eff}\over\partial\varphi_i\partial\varphi_j\partial\varphi_l^*}
\left[{\partial^2 \Sigma_{\rm
eff}\over\partial\varphi_l\partial\varphi_k^*}\right]^{-1} {\partial
\Sigma_{\rm eff}\over\partial\varphi_k} \right](\chi_i\cdot\chi_j)
+2(\chi_t\cdot\chi_t) \right\}+{\mathrm h.c.}
\label{softfer2}
\eea

\vspace{0.3cm}
\noindent
All the expressions of this subsection have been derived from the mechanism
of SUSY breaking discussed in the previous subsection, i.e. an
$S$-dependent $W_{sbr}$ superpotential and
$K(S,S^\dagger)=-M_p^2\log(S+S^\dagger)$. In the alternative no-scale scenario,
with $K(S,S^\dagger)=-3M_p^2\log(S+S^\dagger)$ and
$S$-independent $W_{sbr}$, SUSY is also broken with vanishing
tree-level cosmological constant, as discussed in the previous
subsection. Then the results for the soft breaking terms would be
identical to those obtained in this subsection, with the simple
replacement $\varsigma\rightarrow -2$ in all expressions [compare to
the previous definition in eq.(\ref{varsigma})].

\section{Electroweak Symmetry breaking}
\setcounter{equation}{0}
\renewcommand{\theequation}{5.\arabic{equation}}

The starting point to analyze the electroweak breaking is the complete
Higgs potential, which  can be obtained adding to the SUSY potential
(\ref{Vexp})  the soft terms given by eqs.~(\ref{softsquared}) and
(\ref{softlinear}).  Explicitly, the mass terms read
\bea V_2&=&(|\mu|^2+m_{3/2}^2)(|H_1|^2+|H_2|^2)
-m_{3/2}^2[2(t^2+t^{*2})-5tt^*] \nn\\ &&+\left\{
\tilde{m}_{3/2}^*\left[ 2\lambda \tilde{m}_{3/2}\ -\ \varsigma^*\mu_0\ 
\right]H_1\cdot H_2 +{\mathrm h.c.}  \right\}\ ,
\label{V2}
\eea
In general, the two  contributions to the $\mu$ parameter, $\mu= \mu_0
+\lambda \tilde{m}_{3/2}$, should be considered. If one chooses
$\mu_0=0$ (which is respected by radiative corrections due to SUSY
non-renormalization theorems) one still can get an acceptable
electroweak breaking, with $\lambda \tilde{m}_{3/2}$ as only source
for $\mu$. On the other hand, the option $\lambda=0$ and $\mu=\mu_0$
is problematic for phenomenology as is discussed below. Note also that
the condition $\lambda=0$ is not protected by non-renormalization
theorems so that a non-zero $\lambda$ could be generated by radiative
corrections (provided it is not protected by a Peccei-Quinn symmetry).

To discuss the breaking of the electroweak symmetry we need the rest
of the Higgs effective potential, $V$. Provided $v/\Lambda$ is small,
it is enough to keep in $V$ up to quartic terms in the Higgs fields
(higher order corrections might be included  when necessary). The
cubic terms in the potential are: 
\bea 
V_3&=&3 {m_{3/2}^2\over
\Lambda}(t+t^*)\left[(t+t^*)^2-{9\over 2}tt^*\right]
-{1\over\Lambda}(m_{3/2}^2+2|\mu|^2)(t+t^*)(|H_1|^2+|H_2|^2)\nn\\
&&+\left\{{{\tilde{m}}_{3/2}^*\over\Lambda}\left[
3(\varsigma^*-1)\mu_0 t+(2\mu_0-3\lambda
{\tilde{m}}_{3/2})(t+t^*)\right]H_1\cdot H_2+{\mathrm h.c.}\right\}\ .
\eea 
From this expression we see that  non-zero values for the Higgs
doublets generate a  tadpole for the radion, so that the field $t$
also acquires a VEV as a consequence  of electroweak symmetry
breaking. This VEV is fixed so as to cancel the $t$-tadpole and
represents a small correction [of ${\cal{O}}(v^2/\Lambda^2)$] to the VEV of the
original  radion field $T$. As such it can be absorbed in a small
redefinition of the scale $\Lambda$.

Finally, the quartic terms of the potential are: \bea
V_4&=&{m_{3/2}^2\over \Lambda^2}\left[{17\over 24}(t+t^*)^4-{27\over
8}(t^4+t^{*4})\right]+\left[{m_{3/2}^2\over 2\Lambda^2}
+2{|\mu|^2\over \Lambda^2}  \right] (t+t^*)^2(|H_1|^2+|H_2|^2)\nn\\
&&+\left\{{{\tilde{m}}_{3/2}^*\over2\Lambda^2}\left[
9(1-\varsigma^*)\mu_0 t^2+(5\lambda
{\tilde{m}}_{3/2}-4\mu_0)(t+t^*)^2\right]H_1\cdot H_2+{\mathrm
h.c.}\right\}\nn\\ &&+\left[{g^2\over 8}-{|\mu|^2\over
\Lambda^2}\right](|H_1|^2+|H_2|^2)^2+{{g'}^2\over
8}(|H_1|^2-|H_2|^2)^2\nn\\ &&+\left\{-{g^2\over 2}+
{1\over\Lambda^2}\left[5|\lambda{\tilde{m}}_{3/2}|^2
-3|\mu|^2-2(\lambda{\tilde{m}}_{3/2}\mu^*+{\mathrm h.c.})
\right]\right\}|H_1\cdot H_2|^2 \nn\\
&&+{1\over\Lambda^2}\left[\lambda{\tilde{m}}_{3/2}^*(4\lambda
{\tilde{m}}_{3/2}-5\mu)(H_1\cdot H_2)^2 +{\mathrm h.c.}\right]
\label{V4}
\\
&&+{1\over\Lambda^2}\left[(\lambda m_{3/2}^2-2\mu
{\tilde{m}}_{3/2}^*+3\lambda^2\mu^*{\tilde{m}}_{3/2}-4\lambda|\mu|^2)H_1\cdot
H_2+{\mathrm h.c.}\right](|H_1|^2+|H_2|^2)^2\nn\ .
\eea 
There are significant differences in this potential with respect
to the case of the MSSM (or the NMSSM). It contains new couplings,
both supersymmetric and non-supersymmetric, discussed already in the
two previous sections.  The importance of such terms is manifold, as
we are about to see.  In particular, they turn out to completely
modify the usual pattern  of symmetry breaking, which allows for a
Higgs mass, $M_{h^0}$, much  larger than in the MSSM. If $h^0$ is
discovered, knowledge of its mass ({\it e.g.} if the LEP2 excess is
confirmed and $M_{h^0}\sim\ 115.6$ GeV \cite{LEPSM}) would give
information on $\mu_0/\Lambda$ or $m_{3/2}/\Lambda$.

Another important difference from the MSSM is that the couplings and
masses in $V(=V_2+V_3+V_4)$ as written above are assumed to be
evaluated at the scale $\Lambda$, which is close to the electroweak
scale. As a result, the evolution of these parameters from $\Lambda$
down to the typical scale of SUSY masses ($\sim m_{3/2}$, the relevant
scale for the study of electroweak symmetry breaking) gives in general
a small correction.  This means that the breaking of the electroweak
symmetry is not a radiative  effect, like in the MSSM, but has to
occur at tree-level\footnote{Incidentally, for the very same reason
we do not expect sizeable radiatively-generated off-diagonal entries in
the squark and slepton mass matrices. Hence, these models are very safe
regarding FCNC efects.}.

In the MSSM, with mass terms $V_2=m_1^2 |H_1|^2+ m_2^2|H_2|^2+
m_{12}^2\left( H_1\cdot H_2 +{\mathrm h.c.}\right)$, a proper
electroweak breaking requires, first, that the origin is destabilized,
or
\be m_1^2 m_2^2-m_{12}^4<0\ ,
\label{destab}
\ee
and second, that the potential is not unbounded from below along the
$D$-flat direction $H_1^0=H_2^0$, which requires a positive mass term
along such direction:
\be m_1^2+m_2^2+2m_{12}^2>0\ .
\label{bounded}
\ee
In the MSSM, before radiative corrections, conditions (\ref{destab},
\ref{bounded}) are incompatible if the Higgs masses, $m_1^2$ and $m_2^2$, 
are degenerate (as
it happens here). However, in our case, condition (\ref{bounded}) is
no longer necessary because now there is a non-zero quartic coupling
along the  direction $H_1^0=H_2^0$, which is no longer flat.

Looking at the neutral components of the Higgs doublets,
$H_{j}^0\equiv (h_j^{0r}+ih_j^{0i})/\sqrt{2}$, with $j=1,2$,  the
minimization conditions read
\be {\partial V\over \partial h_1^{0r}}=0\;  ,\;\;\;\;\;\;\;\;
{\partial V\over \partial h_2^{0r}}=0\;  , \ee
from which we can get the Higgs VEVs, or alternatively $v^2\equiv \langle
h_1^{0r}\rangle^2+\langle h_2^{0r}\rangle^2$ and $\tan\beta\equiv
\langle h_2^{0r}\rangle/ \langle h_1^{0r}\rangle$, as functions of the
parameters in $V$. Assuming for simplicity that  all the parameters in
$V$ are real, we obtain, for the VEV $v$:
\be v^2={-m_\beta^2\over \lambda_\beta}\ ,
\label{v}
\ee
with
\bea m_\beta^2&\equiv& (\mu^2+m_{3/2}^2)+ \left[\lambda (2+\varsigma)
m_{3/2}^2 - \varsigma \mu m_{3/2}\right]\sin2\beta\ ,
\label{mbeta}\\
\lambda_\beta&\equiv&{1\over 8}(g^2+{g'}^2)\cos^22\beta
-{\mu^2\over\Lambda^2}\left[1+{3\over 4}
\sin^22\beta+4\lambda\sin2\beta\right]\\
&&+{m_{3/2}^2\over\Lambda^2}\left[{13\over 4}\lambda^2\sin^22\beta
+\lambda\sin2\beta\right] +{\mu m_{3/2}\over\Lambda^2}\left[-{7\over
2} \lambda\sin^22\beta+(3\lambda^2-2)\sin2\beta \right]\nn\ ,
\label{lbeta}
\eea
and, for $\tan\beta$: 
\bea 0&=&\cos
2\beta\left\{4\left[\lambda(v^2+2(2+\varsigma)\Lambda^2)m_{3/2}^2
+[(3\lambda^2-2)v^2-2\varsigma\Lambda^2]\mu m_{3/2}-4\lambda v^2\mu^2
\right]\right.\nn\\
&&\left.-v^2\left[(g^2+{g'}^2)\Lambda^2-26\lambda^2m_{3/2}^2
+28\lambda \mu m_{3/2}+6\mu^2\right] \sin2\beta\right\}\ .
\label{tbeta}
\eea 
The interpretation of (\ref{v}) is straightforward: in the
direction of field space, $\varphi\equiv h_1^{0r}
\cos\beta+h_2^{0r}\sin\beta$, along which the minimum lies (by
definition of $\tan\beta$), the potential is simply 
\be
V(\varphi)={1\over 2}m_{\beta}^2\varphi^2+{1\over 4}\lambda_\beta
\varphi^4\ ,
\label{Valong}
\ee and minimization of this potential leads directly to
eq.~(\ref{v}).  A correct breaking clearly requires \bea
m_{\beta}^2<0\ , \label{cond1}\\ \lambda_\beta >0\label{cond2}\ .
\eea Condition (\ref{cond1}) is equivalent to the MSSM
eq.~(\ref{destab}), while condition (\ref{cond2}) replaces
eq.~(\ref{bounded}).  Before analyzing whether these two conditions
can be satisfied,  we turn to eq.~(\ref{tbeta}), from which we can
determine $\tan\beta$. This equation can be satisfied either by  \be
\tan\beta=1\ , \ee or by \be
\sin2\beta={4\left[\lambda(v^2+2(2+\varsigma)\Lambda^2)m_{3/2}^2
+[(3\lambda^2-2)v^2-2\varsigma\Lambda^2]\mu m_{3/2}-4\lambda v^2\mu^2
\right]\over v^2\left[(g^2+{g'}^2)\Lambda^2-26\lambda^2m_{3/2}^2
+28\lambda \mu m_{3/2}+6\mu^2\right]}\ .
\label{wrongbeta}
\ee This second solution, however, cannot be accepted because it leads
to  problems. To get $\sin2\beta<1$ there must be a cancellation in
the numerator of (\ref{wrongbeta}), which is otherwise of order
$\sim\Lambda^4$ (taking $m_{3/2}$ and $\mu$ not far from $\Lambda$)
while the denominator is only of order $\sim\Lambda^2v^2$. The
cancellation must be  $\lambda (2+\varsigma) m_{3/2}^2 - \varsigma \mu
m_{3/2}\sim {\cal{O}}(v^2)$, but, looking at (\ref{mbeta}), it is
clear that this makes the condition $m_\beta^2<0$ impossible to
satisfy.  Another, more indirect, way of seeing that solution
(\ref{wrongbeta}) is not acceptable is that it leads to a tachyonic
charged Higgs with mass (squared),  $M_{H^{\pm}}^2=-{1\over
4}{g'}^2v^2$.

We are therefore forced to choose $\tan\beta=1$ and, using this in
(\ref{v}) and (\ref{lbeta}), the requirements in
(\ref{cond1}, \ref{cond2}) read 
\bea
\label{vone}
\hspace{-0.75cm} 
m_{\beta=\pi/4}^2&=&(\mu^2+m_{3/2}^2)+ \left[\lambda (2+\varsigma)
m_{3/2}^2 - \varsigma \mu m_{3/2}\right]<0\ ,\vspace*{0.5cm}\\
\hspace{-0.75cm} 
\lambda_{\beta=\pi/4}&=&\left({13\over 4}\lambda^2
+\lambda\right){m_{3/2}^2\over\Lambda^2}+\left(3\lambda^2-{7\over
2}\lambda-2\right){\mu  m_{3/2}\over\Lambda^2}-\left(4\lambda+{7\over
4}\right){\mu^2\over\Lambda^2} >0\ .
\label{lbetaone}
\eea 
We see that the coupling $\lambda$ plays a crucial role in
electroweak symmetry breaking: if $\lambda=0$, then $m_\beta^2<0$ is
easy to satisfy, provided $(2+\varsigma)\mu m_{3/2}>(\mu+m_{3/2})^2\geq 0$,
which requires $\mu m_{3/2}>0$. In that case, we obtain
$\lambda_\beta=-[2\mu m_{3/2}-7\mu^2 /4]/\Lambda^2<0$, which
contradicts the requirement (\ref{cond2}).  Instead, by choosing
$\lambda\neq 0$ conveniently, the two conditions can be satisfied
simultaneously. In order to get $v^2=-m_\beta^2/ \lambda_\beta$
$\sim(246\ {\mathrm GeV})^2$, with $\lambda_\beta$ perturbative,
there must be a cancellation in (\ref{vone}) so as to give
$m_\beta\sim  {\cal{O}}(v)$ [or, more precisely $m_\beta\sim
{\cal{O}}(\lambda_\beta^{1/2}v)$].  This cancellation
requires a tuning similar to that in the MSSM (or even
milder, since $\lambda_\beta$ is normally bigger than in the MSSM). As
we are interested in a situation with $v/\Lambda\ll 1$,  this
cancellation requires
\be \lambda\simeq {-1\over(2+\varsigma)}\left[1-\varsigma{\mu\over
m_{3/2}} +{\mu^2\over m_{3/2}^2}
\right]+{\cal{O}}\left({v^2\over\Lambda^2}\right)\ .  \ee
This is a parabola in the variable $\mu/m_{3/2}$. After substitution
of this expression in  (\ref{lbetaone}), $\lambda_\beta$  is a
polynomial of  fifth degree in $\mu/m_{3/2}$. Therefore, it is
guaranteed that $\lambda_\beta$ takes perturbatively small and
positive values at least near one root of that polynomial.

The reason why $\tan\beta=1$ is a solution of the minimization conditions 
is related to the fact that
the  potential for the neutral Higgs components is symmetric under
$H_1^0\leftrightarrow H_2^0$. In fact, for $g'=0$ the potential  has a
larger symmetry because it is a function of $H_1$ and $H_2$  only
through the combinations
\bea {\mathrm Tr}\left[{\bf H}^\dagger{\bf
H}\right]&=&|H_1|^2+|H_2|^2\ ,\\ {\mathrm Det}\left[{\bf
H}\right]&=&H_1\cdot H_2\ , \eea
where ${\bf H}$ is defined in (\ref{bidoublet}). This means that $V$
is approximately invariant under the $SU(2)_L\times SU(2)_R$ symmetry
discussed in connection to $\Delta\rho$ in section~3.  
For general $\tan\beta$, the spontaneous
breaking of this symmetry leads to three Goldstone bosons:
\bea
G^0&\equiv &{\rm Im}(H_1^0)\cos\beta-{\rm Im}(H_2^0)\sin\beta\ ,\nn\\
\label{pseudoG}
G^+&\equiv &  H_1^+\cos\beta - H_2^+ \sin\beta\ , \\
P^+&\equiv &  H_1^+\sin\beta - H_2^+ \cos\beta\ .\nn
\eea
Here, $G^+$ and $G^0$ are the true Goldstone bosons eaten up by $W^+$
and $Z^0$, while $P^+$ is a pseudo-Goldstone.
Non-zero $g'$ introduces a small
explicit breaking of the symmetry and $P^+$ acquires a
mass
\be M_{P^+}^2=\langle P^-|{\bf M_{\mathrm{ch}}}^2|P^+\rangle=-{1\over
4}{g'}^2v^2\cos^22\beta \ ,
\label{tachyon2}
\ee
(where ${\bf M_{\mathrm{ch}}}^2$ is the charged Higgs mass
matrix). Although $P^+$ is not an eigenstate of ${\bf M_{\mathrm{ch}}}^2$,
$M_{P^+}^2<0$ implies a negative mass eigenvalue, which corresponds to
the charged Higgs mass $M_{H^{\pm}}^2=-{1\over
4}{g'}^2v^2$ mentioned before. This means that $\tan\beta\neq 1$ is 
not a true minimum of the potential.
This problem disappears for $\tan\beta=1$, case
in which $P^+\equiv G^+$, $M_{P^+}^2=0$ and the charged Higgs mass is
not even suppressed by $g'$. 

In general, besides the $g'$-terms,
Yukawa couplings also break the custodial symmetry. As a consequence,
$\tan\beta=1$ is not  protected from radiative corrections (like those
arising from the short running between $\Lambda$ and $m_{3/2}$) and we
expect some small deviations from $\tan\beta=1$.  With this additional
breaking of the custodial symmetry, the mass of the pseudogoldstone
boson [eq.~(\ref{tachyon2})] receives new contributions and 
can lead to $M_{P^+}^2>0$, so that $\tan\beta$ close to (but different from) 1
is not problematic.

Hence, we consider $\tan\beta\simeq 1$ as a quite 
robust prediction of SUSY warped scenarios with universal Higgs masses.
Actually, for other choices of the K\"ahler potential we also expect
$\tan\beta\simeq 1$, as it is explicitly illustrated in Appendix~B.

\section{Higgs sector}
\setcounter{equation}{0}
\renewcommand{\theequation}{6.\arabic{equation}}

In the MSSM,
$\tan\beta=1$ corresponds to a $D$-flat direction in the tree-level
Higgs potential. The tree-level Higgs mass is zero along that
direction and, although radiative corrections can give a  large
contribution to this mass, the experimental limit from LEP2  is
able to exclude the interval $0.5\simlt\tan\beta\simlt 2.4$ at $95\%$
C.L.  \cite{LEPMSSM}. In our model the situation is completely
different because, in addition to the $D$-term contribution to
$\lambda_\beta$ (which is still zero for $\beta=\pi/4$) there are new
corrections which can make $\lambda_\beta>0$ already at tree-level. As
we show later on, the Higgs mass can easily evade LEP2 limits.

After electroweak symmetry breaking, out of the ten initial degrees of
freedom  (d.o.f.'s) in $H_1$, $H_2$ and $t$, three d.o.f.'s [$G^\pm$ and $G^0$
in (\ref{pseudoG})] are
absorbed in the longitudinal  components of the massive gauge bosons,
$W^\pm$ and $Z^0$. The mass eigenstates corresponding to the 
seven remaining  d.o.f.'s  appear as three
$\cal{CP}$-even scalars ($h^0,{h'}^0,H^0$), two $\cal{CP}$-odd
pseudoscalars ($A^0,{A'}^0$) and one charged Higgs ($H^\pm$).   The
calculation of the masses of these physical Higgses requires two
steps. First, one obtains the corresponding mass matrices as the
second derivatives of the effective potential evaluated at the
electroweak minimum. Then one must take into account that a
non-minimal K\"ahler potential leads, after symmetry breaking,  to
non-canonical kinetic terms for scalar fields (see the discussion in
subsection~3.2).  The kinetic terms can be recast in canonical form by
a suitable (non-unitary) field redefinition and the mass matrices have to be
re-expressed in the basis of the canonically-normalized fields. The
final masses are the eigenvalues of these transformed mass
matrices. The details of this procedure are given in Appendix~C. Here
we summarize the results.

In the $\cal{CP}$-even sector, using  
$\{\varphi^0\equiv(h_1^{0r}+h_2^{0r})/\sqrt{2}, 
H^0\equiv(h_1^{0r}-h_2^{0r})/\sqrt{2}, t_R^0\}$,
(with $h_{1,2}^{0r}\equiv \sqrt{2} {\rm Re}[H_{1,2}^0]$ and $t_R^0\equiv \sqrt{2} 
{\rm Re}[t]$) as a convenient basis, we find that $H^0$ is a mass-eigenstate,
with $M_{H^0}^2$ controlled by $\mu m_{3/2}$ and $m_{3/2}^2$ [see (\ref{MHN})], 
while $\varphi^0$ and $t_R^0$ are in
general mixed to give the two mass eigenstates $h^0$ and ${h'}^0$. 
The exact masses of $h^0$ and ${h'}^0$ can be simply obtained as 
the eigenvalues of a $2\times 2$ matrix, but the resulting expressions are not very
illuminating. In this respect it is useful to note that the mass of the lightest
$\cal{CP}$-even Higgs, $h^0$, satisfies the mass bound (${\bf M_{h}^2}$ is the
$\cal{CP}$-even mass matrix)
\bea
M_{h^0}^2&\leq&\langle \varphi^0| {\bf M_{h}^2}|\varphi^0\rangle=
2(\lambda_\beta+\xi_\beta) v^2\ ,
\label{bound1}\\
M_{h^0}^2&\leq&\langle t_R^0 | {\bf M_{h}^2}| t_R^0 \rangle\ ,
\label{bound2}
\eea
with $\lambda_\beta$ defined in (\ref{lbeta}) and $\xi_\beta$, which 
represents  a shift in the quartic coupling $\lambda_\beta$  due to kinetic mixing 
between $H_{1,2}^0$ and $t$, is given in Appendix~C. 
The virtue of this bound, eq.~(\ref{bound1}), is that it is controlled
by the electroweak scale $v$: it can only be made heavy at the expense of
making the coupling $(\lambda_\beta+\xi_\beta)$ strong. 
The existence of such bound follows from 
general arguments (see {\it e.g.} \cite{bound} and references therein).
On the other hand,
as we explicitly show below, the lightest $\cal{CP}$-even Higgs boson has 
in general some singlet component coming from $t_R^0$ and this 
makes the couplings of $h^0$ different from the SM ones. 

In the  $\cal{CP}$-odd sector, we use the basis $\{G^0\equiv (h_1^{0i}-h_2^{0i})/\sqrt{2}, 
a^0\equiv(h_1^{0i}+h_2^{0i})/\sqrt{2}, t_I^0\}$ (where  
$h_{1,2}^{0i}\equiv \sqrt{2}{\rm Im}[H_{1,2}^0]$ and $t_I^0\equiv \sqrt{2} 
{Im}[t]$). The neutral Goldstone boson, $G^0$, is an exact eigenstate, with zero mass,
while in general $a^0$ and $t_I^0$ are mixed  to give the two mass eigenstates
$A^0$ and ${A'}^0$. From the expressions in Appendix~C one 
can see that, for $\mu$ and $m_{3/2}$ moderately larger than $v$, $a^0$ 
and $t_I^0$ are approximate mass eigenstates up to ${\cal{O}}(v/\Lambda)$ corrections:
$A^0\simeq a^0$ and ${A'}^0\simeq t_I^0$. 

In the charged Higgs sector we find that the charged Goldstone boson is 
$G^+\equiv (H_1^{+}-H_2^+)/\sqrt{2}$ and the physical Higgs $H^+\equiv (H_1^{+}+H_2^+)/\sqrt{2}$.
The mass for this charged Higgs boson is
\be
M_{H^+}^2=M_{H^0}^2-{1\over 4}{g'}^2v^2\ ,
\label{massdeg}
\ee
with $M_{H^0}^2$ as defined in eq.~(\ref{MHN}).

The pattern of Higgs masses just described conforms neatly with the
requirements of the underlying symmetries of the Higgs potential, $V$.
We have already shown that, after electroweak
symmetry breaking with $\tan\beta=1$, $V$ still has an approximate
$SU(2)_{L+R}$ custodial symmetry  only broken by small $g'$-terms.
Therefore, the physical Higgs fields belong in 
representations of $SU(2)_{L+R}$, with the members of the same multiplet
having degenerate masses [up to small ${\cal{O}}({g'}^2v^2)$ splittings].
To show how this happens, notice first that a real $SU(2)_{L+R}$ 
triplet, $\{\xi^+,\xi^0,\xi^-\}$, can be written as the $2\times 2$ matrix
\be
{\bf T_\xi}={1\over \sqrt{2}}{\bma\sigma}^a\xi_a=
\left[
\begin{array}{cc}
\xi^{0}/\sqrt{2} & \xi^+ \\
\xi^- & -\xi^{0}/\sqrt{2}
\end{array}\right]\ ,
\label{adjoint}
\ee
with $\xi^0\equiv \xi_3$ and $\xi^\pm\equiv (\xi_1\mp i\xi_2)/\sqrt{2}$.
The transformation law  ${\bf T_\xi}\rightarrow {\bf U_{L+R}} {\bf
T_\xi}{\bf U_{L+R}}^\dagger$ holds, since (\ref{adjoint}) is clearly in the
adjoint representation. Using now the bi-doublets defined in (\ref{bidoublet})
we can make manifest the $SU(2)_{L+R}$ representations of the Higgs 
fields, by writing
\bea
{\bf H}+\overline{\bf H}&=&\varphi^0 {\bf I}_2+i\sqrt{2}\left[
\begin{array}{cc}
G^{0}/\sqrt{2} & G^+ \\
G^- & -G^{0}/\sqrt{2}
\end{array}\right]\equiv \varphi^0 {\bf I}_2+i\sqrt{2} {\bf T_G}\ ,\\
-i({\bf H}-\overline{\bf H})&=&a^0 {\bf I}_2-i\sqrt{2}\left[
\begin{array}{cc}
H^{0}/\sqrt{2} & H^+ \\    
H^- & -H^{0}/\sqrt{2}    
\end{array}\right]\equiv a^0 {\bf I}_2-i\sqrt{2} {\bf T_H}\ ,
\eea
corresponding to the decomposition under $SU(2)_{L+R}$ of a
$SU(2)_{L}\times SU(2)_{R}$ bi-doublet:  
$(2,2)\sim 2\otimes 2=1\oplus 3$.
These formulae show that $\varphi^0$ and $a^0$ are $SU(2)_{L+R}$ singlets
[and this is consistent with the fact that there can be $\varphi^0-t_R^0$
and $a^0-t_I^0$ mixing; $t$ being obviously a $SU(2)_{L+R}$ singlet too]. 
Then the three
Goldstone bosons form a real $SU(2)_{L+R}$ triplet, ${\bf T_G}$ and have
zero masses. Finally, $H^\pm$ and $H^0$ form another triplet, ${\bf T_H}$.
As a consequence, $H^0$ does not mix with other neutral scalars and is
mass-degenerate with $H^\pm$, up to ${\cal{O}}({g'}^2v^2)$ corrections
[see eq.~(\ref{massdeg})].
 
\begin{figure}[t]
\centerline{
\psfig{figure=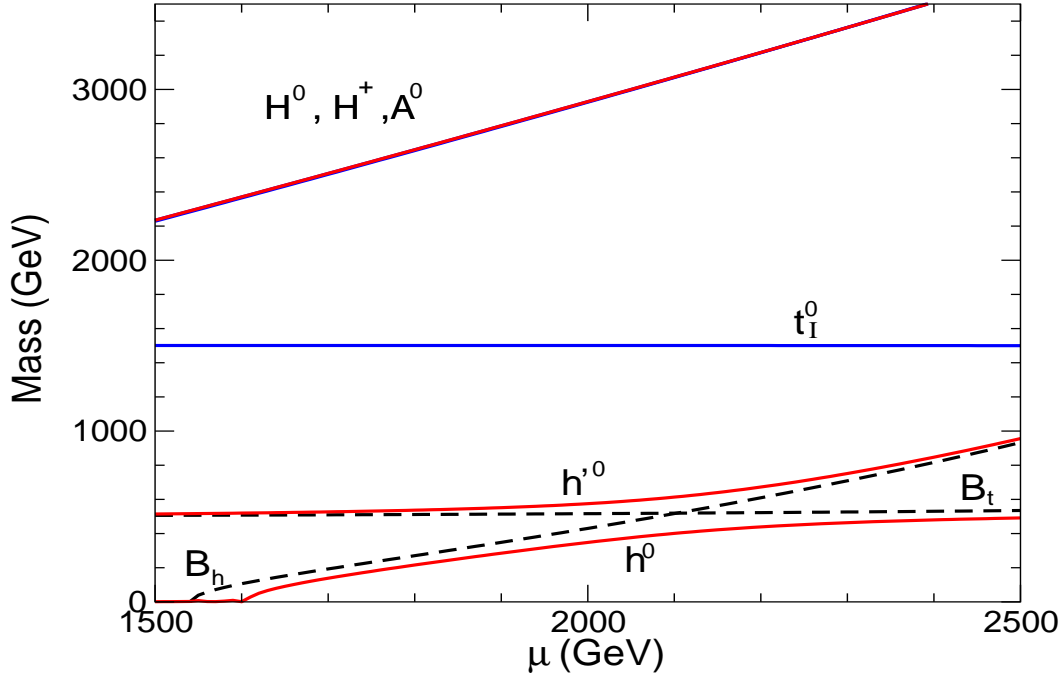,height=10cm,width=10cm,angle=-90,bbllx=2.cm,%
bblly=5.cm,bburx=21.cm,bbury=22.cm}}
\caption
{\footnotesize
Higgs boson masses, in GeV, as functions of $\mu$
for $\Lambda=4$ TeV, $m_{3/2}=500$ GeV and $\varsigma=1+\sqrt{3}$.
}
\end{figure}

Note also that, if the $v$-independent part of $M_{H^0}^2\simeq M_{H^\pm}^2$
and $M_{A^0}^2$ dominates over the $v$-dependent part (and this is
the case of interest to us, with $\mu$ and $m_{3/2}$ larger than $v$, 
which is much smaller than $\Lambda$), then $H^\pm$, $H^0$ and $A^0$
arrange in a heavy nearly-degenerate $SU(2)_L$ Higgs doublet
\be
H={1\over\sqrt{2}}\left[H_2 -\overline{H}_1\right]=
\left[
\begin{array}{c}
H^+\\
{1\over\sqrt{2}}(H^0+iA^0)
\end{array}
\right]\ ,
\ee
which does not participate in the breaking of the electroweak 
symmetry\footnote{The combination relevant for electroweak  
breaking is the orthogonal one, $\left[
\begin{array}{c}
G^+\\
{1\over\sqrt{2}}(h^0+iG^0)
\end{array}
\right]$.}
and has mass (squared) 
$M_H^2\sim 2\left[\varsigma\mu m_{3/2}-\lambda(2+\varsigma)m_{3/2}^2 \right]$. 
In addition, there is another heavy pseudoscalar, ${A'}^0\simeq t_I^0$, with 
mass  $\sim 
3m_{3/2}$ 
and two scalars, $\varphi^0$ and $t_R^0$,
which can mix and be moderately light (the mass eigenstates are 
$h^0$, ${h'}^0$).
This behaviour is clearly
shown by the numerical example given in figure~1. We take $\Lambda=4$ TeV, 
$m_{3/2}=500$ GeV, $\varsigma=1+\sqrt{3}$ and let $\mu$ vary inside the range 
that gives a proper electroweak breaking. For each value of $\mu$, 
$\lambda$ is chosen so as to satisfy $v^2=(246 {\mathrm\ GeV})^2$. In the 
region shown $\lambda$ varies from $-0.4$ to $-3$.  We
present the Higgs masses as a function of $\mu$: 
$H^0$, $A^0$ and $H^+$ are heavy and appear nearly-degenerate 
in the upper part of the plot;
the other curves give the masses for the two light
${\cal{CP}}$-even scalars, $h^0$, ${h'}^0$ and the second ${\cal{CP}}$-odd
scalar, which we label $t_I^0$ because it is basically a pure $t_I^0$
state.  We also show the two upper bounds on $M_{h^0}$ as given by
eq.~(\ref{bound1}) (labelled $B_h$) and eq.~(\ref{bound2}) (labelled
$B_t$).
\begin{figure}[t]
\centerline{
\psfig{figure=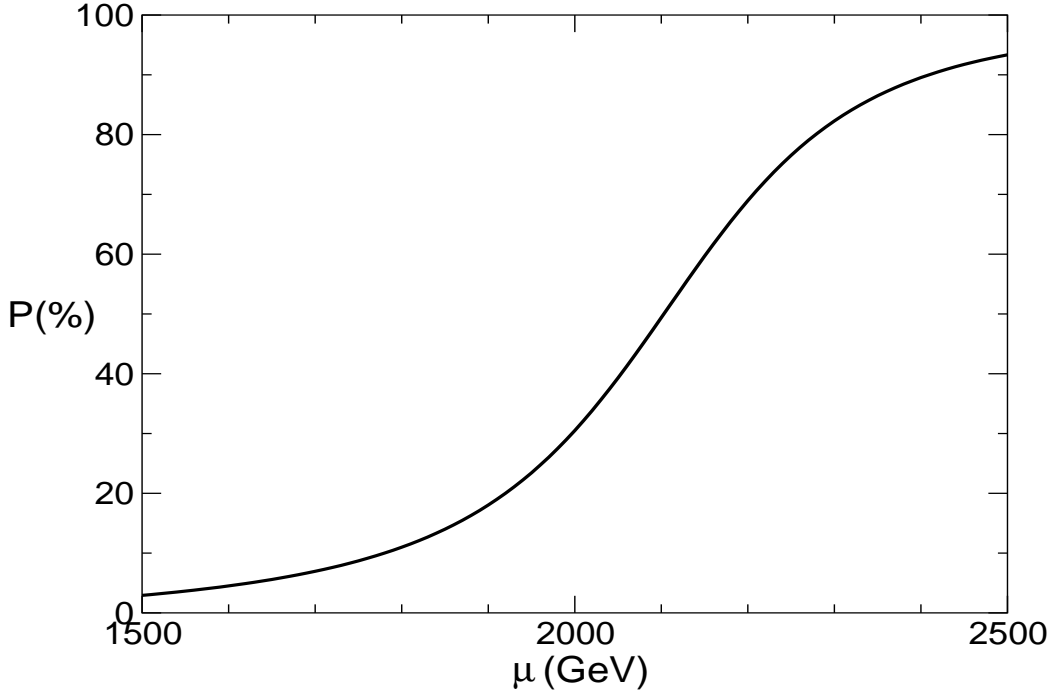,height=10cm,width=10cm,angle=-90,bbllx=2.cm,%
bblly=5.cm,bburx=21.cm,bbury=22.cm}}
\caption
{\footnotesize
Percentage, $P$, of singlet ($t_R^0$) component in the lightest Higgs 
$h^0$. The percentage of singlet component in the second-to-lightest 
Higgs, ${h'}^0$, is $100\%-P$.}
\end{figure}

From figure~1 we see that the lightest Higgs boson, $h^0$, can be quite
heavy compared with the MSSM case (up to $M_{h^0}\sim m_{3/2}= 500$ 
GeV in this particular example). To ascertain the properties of this light 
Higgs it is necessary to study its composition in terms of $\varphi^0$ 
[which belongs to a $SU(2)_L$-doublet] and $t_R^0$
(which is a gauge singlet). This is shown in figure~2, which gives the
percentage of singlet component in $h^0$ (that of ${h'}^0$ is exactly
complementary to it). In the region of low values of $\mu$, when
$M_{h^0}$ approaches $B_h$, $h^0$ is mostly $\varphi^0$ ({\it i.e.}
SM-like), with a small but non-negligible  ($\sim 5\%$) singlet 
component. For larger values
of $\mu$, when $M_{h^0}$ approaches $B_t$, $h^0$ is mostly $t_R^0$,
with a large ($\sim 95\%$) singlet component.  In this 
respect, the true Higgs in this latter region is ${h'}^0$, which is
 significantly heavier than $h^0$. In fact, one
should worry about perturbativity if a SM-like Higgs gets as heavy as
${h'}^0$ does in that region. If we demand that the quartic coupling
$(\lambda_\beta+\xi_\beta)$ remains below $\sim 4$ (see {\it e.g.}
\cite{pert}), we obtain an upper limit $M_{{h'}^0}\simlt 700$ GeV, similar
to the usual perturbativity limit on the SM-like Higgs boson. (This sets also
an upper limit $\sim 2250$ GeV on the value of $\mu$.) The fact that the Higgs
upper 
bound is so much higher than the usual bound $M_h^0\simlt 205$ GeV
in general SUSY models \cite{EQ} is due to the smallness of the cut-off scale $\Lambda$
(the usual bound assumes the model to be valid all the way up to the Planck scale).

So far we have discussed only tree-level masses. One 
should add to them radiative corrections, which
are expected to be sizeable in general, because the stops have large masses in 
this model. These top-stop radiative corrections have the same form 
as in the MSSM \cite{MSSMrad} and can raise the Higgs mass significantly.

\section{Neutralino sector}
\setcounter{equation}{0}
\renewcommand{\theequation}{7.\arabic{equation}}

In this section we present the $5\times 5$ neutralino mass matrix
after electroweak symmetry breaking. First there are 
supersymmetric contributions to Higgsino and radino masses from
eq.~(\ref{Yuk}). To this one should
add gaugino soft masses [$M_1$ and $M_2$ for $U(1)_Y$
and $SU(2)_L$, respectively] and the soft-terms of eq.~(\ref{softfer2})
which, more explicitly, give the mass terms:
\be
\delta{\cal{L}}_{\rm soft}={\tilde{m}}_{3/2}\left[(\chi_t\cdot\chi_t)
-\lambda(\chi_{H_1^0}\cdot\chi_{H_2^0})
\right]+{\mathrm h.c.}
\ee
The last step to find the neutralino mass matrix is to normalize the
neutralino fields so as to get canonical kinetic terms. This is exactly
similar to what was done for the Higgs bosons in Appendix~C, and in fact
the field-redefinition of the Higgsinos is identical.
With all these ingredients, the neutralino mass matrix, in the basis
$\{\lambda_B,\lambda_{W_3},\chi_{H_1^0},\chi_{H_2^0},\chi_t\}$
and neglecting ${\cal{O}}(v^2/\Lambda)$ terms, reads:
\be
{\bf {\cal{M}}}_{\tilde{\chi}^0}\simeq\left[
\begin{array}{ccccc}
M_1 & 0   & -M_Zs_wc_\beta &  M_Zs_ws_\beta & 0 \vspace{0.1cm}\\
0   & M_2 &  M_Zc_wc_\beta & -M_Zc_ws_\beta & 0 \vspace{0.1cm}\\
-M_Zs_wc_\beta & M_Zc_wc_\beta  & 0 & -\mu  &
{\mu}_{1}{\displaystyle{v\over\Lambda}} \vspace{0.1cm} \\
M_Zs_ws_\beta & -M_Zc_ws_\beta  &
-\mu & 0 & {\mu}_{2}{\displaystyle{v\over\Lambda}} \vspace{0.1cm} \\
0 & 0 & {\mu}_{1}{\displaystyle{v\over\Lambda}}  &
 {\mu}_{2}{\displaystyle{v\over\Lambda}}  &
2{\tilde{m}}_{3/2} \end{array}\right]\ ,
\label{neutralinomatrix}
\ee
where $\theta_w$ is the Weinberg angle; 
we use the short-hand notation $s_\beta=\sin\beta$,
$c_w=\cos\theta_w$, etc; and
\bea
{\mu}_1&\equiv & {1\over 4}(1+\lambda)(2{\tilde{m}}_{3/2}-\mu)+\sqrt{2}\mu_0\sin\beta\ ,\nn\\
{\mu}_2&\equiv &{1\over 4}(1+\lambda)(2{\tilde{m}}_{3/2}-\mu)+\sqrt{2}\mu_0\cos\beta\ .
\eea
We note that $\mu_0$ has combined with $\lambda {\tilde{m}}_{3/2}$ to give $\mu$
as usual (the same happens in the chargino mass matrix, which is
of standard form), and $\chi_t$ receives a soft mass $2{\tilde{m}}_{3/2}$. This 
solves a possible problem with a very light neutralino, because the SUSY
mass of $\chi_t$ is only of order $\mu_0 v^2/\Lambda^2$.
Notice also in (\ref{neutralinomatrix}) that the gauginos do not have a direct 
mixing with the radino, the reason being that the latter is a gauge singlet.
The structure of the neutralino mass matrix (\ref{neutralinomatrix}) is different from that
of the NMSSM (see {\it e.g.} \cite{Zerwas} and references therein). Once
again the 
difference can be traced back to the non-minimal K\"ahler potential. 

\section{Summary and conclusions}

Supersymmetrization of warped scenarios is helpful to avoid some 
shortcomings of the original Randall-Sundrum construction. In particular,
SUSY helps to explain the correlations between the
brane tensions and the bulk cosmological constant, and 
to protect the hierarchy against destabilization by radiative 
corrections. In addition, SUSY is likely a necessary ingredient to make contact 
with warped superstring constructions.

The main goal of this paper has been  to derive the low-energy
effective theories of supersymmetric warped constructions and to
extract from  them the low-energy phenomenology. They turn out to be
pretty unconventional, which might be useful to alleviate some of the
drawbacks of the MSSM. An important example is the upper bound  on the
Higgs mass: while it is starting to be worrisome in the conventional
MSSM, we have showed that it completely disappears here.  On the
other hand, these  constructions offer several characteristic
phenomenological features,  which depart from those of ordinary
supersymmetric scenarios.

Next we summarize the main results and conclusions of the paper.

\subsubsection*{Effective supersymmetric theory}

As discussed in sect.2, the most distinctive feature of warped
supersymmetric constructions with respect to ordinary supersymmetric
versions of the SM is the fairly model-independent fashion in which
the radion couples to the visible fields, {\em both} in the
superpotential and the K\"ahler potential. These are given by eq.(\ref{EffK})
[or alternatively eq.(\ref{EffK2})] and eq.(\ref{EffW}). The first equation
still leaves room for some arbitrariness in
the form of the K\"ahler potential, $K$. For practical computations we have used 
the form of $K$ inspired by ref.~\cite{Pok}, which corresponds to 
eq.(\ref{ansatzSigma}).
For other ans\"atze, most of the basic results hold, as they are a consequence 
of the above-mentioned model-independent features. This is briefly illustrated 
in Appendix~B for an alternative, and somewhat simpler, choice of $K$.

In our scheme, all the visible fields, including the ordinary Yang-Mills
fields, live in the visible brane. We have included, however, a
universal hypermultiplet in the bulk, which plays the role of the
usual moduli fields (in particular the dilaton), abundant in
superstring constructions. 
As discussed in sect.3, we have not made any
assumption about the form  of the gauge kinetic functions, $f_A$.
They  are not very important for the phenomenology discussed 
here, except for the predictions on gaugino masses.

The most important novelty of the effective supersymmetric theory is
that the K\"ahler potential is not minimal 
in the $M_p\rightarrow\infty$ limit.
This is illustrated by eq.(\ref{Keff})
in our particular scenario.  As a result the low-energy 
globally-supersymmetric theory (whose general Lagrangian is given in
Appendix~A) shows distinctive features. In particular,
the radion mixes with the Higgs fields (and the radino
with the Higgsinos)  through kinetic and  mass terms after electroweak
symmetry breaking. Also, there arise new supersymmetric couplings (or
corrections to ordinary couplings),  suppressed only by
$\mu_0/\Lambda$ (or $m_{3/2}/\Lambda$ after SUSY breaking), where
$\Lambda ={\cal{O}}$(TeV) is the typical mass scale in the visible
brane, given by eq.(\ref{Tvev}), and $\mu_0$ is the $\mu$-parameter in the
superpotential. [Incidentally, there is no $\mu$-problem in these
scenarios since, besides the known mechanisms to generate it, its
natural value is ${\cal{O}}(\Lambda)$, as any other mass scale in the
visible brane.] These new couplings include extra quartic  couplings
for the Higgses, which completely change the Higgs
phenomenology (a complete analysis, however, requires to take into 
account the SUSY breaking contributions, see below).
Finally, there are higher order operators  ({\it e.g.}
with two derivatives) suppressed only  by inverse powers of
$\Lambda$. These give rise to tree-level contributions to the $\rho$
parameter, setting lower bounds on the value of $\Lambda$, see 
eq.(\ref{ro-bound}).

Finally, let us note that in the  
$\Lambda\rightarrow\infty$ limit, the radion decouples and the superpotential and 
the K\"ahler potential recover their ordinary ``minimal'' forms.

\subsubsection*{Supersymmetry breaking}

SUSY breaking plays an essential role for the stabilization of the
radion field and is mandatory for a correct low-energy phenomenology.
In our scheme SUSY is broken by brane superpotentials 
$W_{sbr}=W_h+e^{-3 T}W_v$
located  at the
two branes. This is similar to the scheme of ref.~\cite{Pok}, but with one
important difference: we allow the superpotentials to depend on the
``dilaton'' field, $S$ (i.e. the bulk moduli). 
Then SUSY {\em can} be
broken at vanishing cosmological constant (this constrains
the form of the $S$-dependence). The $T$ field gets stabilized with a
VEV related to the SUSY breaking scale. Provided the dependence
of $W_{sbr}$ on $S$ is factorizable, $F_T=0$ and the breaking occurs 
entirely along the $F_S$ direction.

The corresponding soft terms 
show universality as a result of the flavour blindness of the $S$-couplings.
They are more or less conventional, except for the appearance of non-standard
trilinear terms of the form $t^*|\varphi|^2+{\mathrm h.c.}$
(where $\varphi$ is a chiral scalar field) 
and dimension 4 (and higher) SUSY-breaking operators suppressed 
only by powers of $m_{3/2}/\Lambda$. We show that they do not 
introduce dangerous quadratic divergences able to spoil the
naturalness of the electroweak breaking. However, they play a relevant 
role in the phenomenology of these scenarios.

\subsubsection*{Electroweak breaking}

Due to the universality of the soft breaking terms (in particular the
Higgs mass terms) and the short range for the RG running,
$\tan\beta\simeq 1$ appears as a typical feature of these models. In
an ordinary MSSM this would be at odds with the present bounds on the
lightest Higgs boson. Here this is not so, since electroweak
breaking takes place in a rather unconventional way. It remains a fact
that proper electroweak breaking requires that the origin of the 
Higgs-field space is destabilized, which implies $m_1^2 m_2^2-m_{12}^4<0$.
However, the usual complementary condition $m_1^2+m_2^2+2m_{12}^2>0$, to 
avoid an unbounded from below ($D$-flat) direction along
$H_1^0=H_2^0$, is not required anymore. This is because, as stated above, 
$V(H)$ contains extra quartic couplings which lift the $D$-flat direction.
A complete expression for the Higgs potential is given 
in eqs.(\ref{V2}--\ref{V4}).

These facts imply that electroweak breaking occurs at tree-level (it
is not a radiative effect, like in the MSSM). Still, the breaking is
completely natural in the sense that the Higgs fields are the only
ones with a $\mu$-term (and the corresponding bilinear soft term), and
therefore (provided the soft breaking masses are positive) are the
only scalar fields of the theory which can be destabilized 
at the origin.

\subsubsection*{Higgs and neutralino sectors}

Due to the unconventional electroweak breaking the usual expressions
for the Higgs spectrum do not hold any longer. This is true, in
particular, for the usual MSSM tree-level bound on the mass of the
lightest Higgs field, $M_{h^0}^2\leq M_Z^2 \cos^2 2\beta$, which is now
replaced by the bounds of eqs.(\ref{bound1}, \ref{bound2}). Actually,
we show that $M_{h^0}$ can be as large as 700 GeV, a size similar
to the usual perturbativity limit on the SM-like Higgs boson. 

The complete spectrum of the Higgs sector is a bit more complicated
than in the MSSM, since two additional real fields (one
$\cal{CP}$-even and one $\cal{CP}$-odd) enter the game, namely the
real and imaginary parts of the radion field.  (The complete
expressions for the tree-level spectrum are given in Appendix~C.)
Actually, another important difference with ordinary cases is that
the neutral Higgs field eigenstates get a sizeable component of the
radion.  In particular, the percentage of radion in the lightest Higgs
field, ${h^0}$, depends on the value of $\mu$, but it is non-negligible 
in all cases. This affects the experimental properties of the 
light Higgs field, since the radion is a singlet.

Likewise, as stated above, the Higgsinos get mixed with the
radino. After the SUSY and electroweak breakdown, this results in a
non-conventional $5\times5$ neutralino mass matrix, which is
explicitly given in eq.(\ref{neutralinomatrix}).

\vspace{0.4cm} 

\noindent
To conclude, the low-energy effective theory from supersymmetric
warped  constructions presents many interesting differences (which are
sometimes advantages) from the ordinary MSSM. They are caused by the
non-minimal (and quite model-independent) form of the superpotential
and the K\"ahler potential, a fact that could be also present in other
models with extra-dimensions where the fundamental scale is
${\cal{O}}$(TeV). This offers novel ways of accommodating and
testing new physics.

\setcounter{equation}{0}
\renewcommand{\theequation}{8.\arabic{equation}}

\section*{Acknowledgments}
We thank A.~Falkowski, S.~Lalak, C. Mu\~noz and R.~Sundrum for very 
useful discussions and correspondence.

\newpage
\section*{A. Lagrangian for general K\"ahler potentials}
\setcounter{equation}{0}
\renewcommand{\theequation}{A.\arabic{equation}}

Given a superpotential $W(\Phi)$ and a
general K\"ahler potential $K(\Phi,\Phi^\dagger)$ 
the corresponding supersymmetric Lagrangian is:
\be
{\mathcal{L}} = {1\over2}\left\{2
\left[W(\Phi)\right]_{\mathcal{F}}+{1\over2}\left[K[\Phi,\Phi^\dagger
\exp(-2{\bf t_A} V_A)]\right]_D -{1\over2} \left[f_{AB}(\Phi)(W_{AL}^T
\epsilon W_{BL})\right]_{\mathcal{F}}\right\}+h.c. 
\label{Lgen}
\ee
Here, ${\bf t_A}$ are Hermitian generators of
the gauge group algebra, labelled by $A$; $V_A$ are the corresponding gauge 
vector superfields; $W_{AL}$ are the chiral  
field-strength spinor superfields; $\epsilon_{\alpha\beta}$ is the 
totally antisymmetric tensor and $f_{AB}(\Phi)$ is the gauge kinetic function.

Assuming further that $W$ and $K$
do not depend on field derivatives, the Lagrangian (\ref{Lgen}) written
in components reads (${\cal{G}}_{ij}\equiv
\partial^2 K/\partial \phi_i\partial \phi_j^*$):
\begin{eqnarray}
{\mathcal{L}} &=& {1\over 2}\left\{
{\cal{G}}_{ij}
\left[i \zeta_j^\dagger {\bma \sigma}^\mu D_{\mu} \zeta_i +
{\mathcal{F}}_i {\mathcal{F}}_j^* + D_\mu \phi_i D^\mu \phi^*_j\right]
-i{\partial^3 K(\phi,\phi^*)\over  \partial \phi_i \partial \phi_j
\partial\phi_l^*}(\zeta_i^T{\bma\sigma}_2 \zeta_j){\mathcal{F}}_l^* \right.\nn\\
&+& i{\partial^3 K(\phi,\phi^*)\over  \partial \phi_i \partial \phi_j
\partial\phi_l^*}(\zeta_l^\dagger {\bma\sigma}^\mu \zeta_j)D_\mu \phi_i
 + {1\over4}{\partial^4 K(\phi,\phi^*)\over  \partial \phi_i \partial \phi_j
\partial\phi_l^* \partial\phi_k^*}(\zeta_i^T{\bma\sigma}_2 \zeta_j)
(\zeta_k^\dagger{\bma\sigma}_2 \zeta_l^*)\nn\\
&-&i{\partial^2W(\phi)\over\partial\phi_i \partial \phi_j} (\zeta_i^T{\bma\sigma}_2 \zeta_j)
+2{\mathcal{F}}_i{\partial W(\phi)\over\partial\phi_i}
- {1\over 4}(\lambda_A^T{\bma\sigma}_2 \lambda_B)
(\zeta_i^T{\bma\sigma}_2 \zeta_j){\partial^2 f_{AB}(\phi) \over \partial
\phi_i \partial \phi_j}\nn\\
&-& {i\over 2}(\lambda_A^T{\bma\sigma}_2 \lambda_B){\mathcal{F}}_i{\partial f_{AB}(\phi) \over \partial
\phi_i} + {\sqrt{2}\over 4}{\partial f_{AB}(\phi)\over \partial
\phi_i}\left[i(\lambda_B^T{\bma\sigma}_2\bar{\bma\sigma}^\mu{\bma\sigma}^\nu \zeta_i)f_{A \mu \nu} - 2 (\lambda_B^T
{\bma\sigma}_2\zeta_i)D_A\right] \nn\\
&+& f_{AB}(\phi)\left[-i\lambda_B^\dagger{\bma\sigma}^\mu D_{\mu} \lambda_A - {1\over 4}F_{A\mu
\nu}F_B^{\mu \nu} +{i\over 8}\epsilon_{\mu \nu \rho \sigma}F_A^{\mu \nu}F_B^{\rho
\sigma} + {1\over 2}D_A D_B \right]\nn\\
&-&\left.{\partial K(\phi,\phi^*)\over\partial\phi_i^*}D_A(\phi^\dagger{\bf t_A})_i+
2\sqrt{2}{\partial^2K(\phi,\phi^*)\over\partial\phi_i\partial\phi_j^*}
({\bf t_A}\phi)_i(\zeta_{j}^\dagger{\bf \sigma}_2\lambda_{A}^*)
\right\} +{\mathrm h.c.}
\label{generaL}
\end{eqnarray}
with $\Phi_i=(\phi_i,\zeta_{i},{\cal{F}}_i)$, $V_A=(V_{A\mu},\lambda_A,D_A)$;
$D_\mu \phi_i\equiv\partial_\mu\phi_i-i({\bf t_A}\phi)_iV_\mu^A$ and a similar 
definition for $D_\mu \zeta_{i}$.
Eq.~(\ref{generaL}) corrects formula
(27.4.42) of \cite{Weinberg} by including the gauge terms of $1/2[K]_D$ which are not  
absorbed in the replacement $\partial_\mu\rightarrow D_\mu$, {\it i.e.} the term with a single
derivative of $K$ and the $K$-dependent term that mixes matter fermions with gauginos 
[they are collected in the last line of (\ref{generaL})].

The auxiliary fields ${\mathcal{F}}_i$ and $D_A$ can be eliminated from ${\mathcal{L}}$
above by using their equations of motion, which give:
\bea
{\mathcal{F}}_i^*&=&
{\cal{G}}_{ij}^{-1}
\left\{-{\partial W(\phi)\over \partial\phi_j}+
{i\over2}{\partial^3 K(\phi,\phi^*)\over\partial\phi_k^*\partial\phi_l^*\partial\phi_j}
(\zeta_l^\dagger{\bma \sigma}_2\zeta_k^*)
+{i\over 2}{{\partial f_{AB}(\phi)\over \partial\phi_j}}
(\lambda_A^T{\bma\sigma}_2\lambda_B)
\right\}\ ,\nn\\
D_A&=&f_{AB}^{-1}(\phi)\left[
{\partial K(\phi,\phi^*)\over\partial\phi_i}({\bma t_B}\phi)_i
+{1\over\sqrt{2}}{\partial f_{BC}(\phi)\over\partial\phi_i}
(\lambda_C^T{\bma\sigma}_2\zeta_i)
\right]\ .
\eea

\newpage

\section*{B. Alternative choice of K\"ahler potential}
\setcounter{equation}{0}
\renewcommand{\theequation}{B.\arabic{equation}}

Throughout the paper we have used the choice (\ref{ansatzSigma}) for the general
expression
of $K$, (\ref{EffK2}), which corresponds to the ansatz given in
ref.~\cite{Pok}. In the 
$M_p\rightarrow\infty$ limit this goes to (\ref{Keff}). In this Appendix we take
$\Phi_{\rm vis}=\sum_i|\hat\varphi_i|^2+(\lambda \hat H_1\cdot \hat H_2+{\mathrm
h.c.})$, to be
plugged in the general expression of $K$, (\ref{EffK}). This is perhaps
closer in spirit to the derivation of $K$ given in \cite{LS}. Then, in the
$M_p\rightarrow\infty$ limit we obtain
\be
\boxeq{
K_{\rm eff}=\Lambda^2 \exp\left[-{t+t^*\over\Lambda}\right]
\left\{
1+{1\over\Lambda^2}\left[\sum_i|\varphi_i|^2+\left(
\lambda H_1\cdot H_2 + {\mathrm h.c.}\right)\right]
\right\}
}
\label{KeffLS}
\ee
We can use the formulae given in the text for generic $\Sigma_{\rm eff}$
if we make the replacement 
\be
\Sigma_{\rm eff}\rightarrow \Lambda^2
\log [1+\Sigma_{\rm eff}/\Lambda^2]\ ,
\label{repla}
\ee
which transforms (\ref{Keff}) into (\ref{KeffLS}). 
The supersymmetric scalar potential is obtained directly from  (\ref{Vgen2})
by making such replacement and working out the inverse matrix. The 
final result is
\bea
V_{\rm SUSY}&=&e^{(t+t^*)/\Lambda}\sum_i\left|{\partial W_{\rm eff}\over
\partial\varphi_i}\right|^2+
{1\over K_{\rm eff}N_{eff}}\left|
{\partial W_{\rm eff}\over \partial\phi_i}
{\partial \Sigma_{\rm eff}\over\partial\varphi_i^*}
-3W_{\rm eff}\right|^2
\nn\\
&+&
{1\over 4}e^{-2(t+t^*)/\Lambda}
\left\{f_{A}^{-1}(\phi)\left[
{\partial \Sigma_{\rm eff}\over\partial\varphi_i}({\bf t_A}\varphi)_i\right]^2+
{\mathrm h.c.}\right\}\ .
\label{Vgen2LS}
\eea
In this equation, $\Sigma_{\rm eff}$ is as given in (\ref{Seff}) and
we have defined
\be
N_{\rm eff}\equiv 1+{\Sigma_{\rm eff}\over\Lambda^2}-{1\over\Lambda^2}
\sum_i\left|{\partial\Sigma_{\rm eff}\over\partial\varphi_i}\right|^2
= 1-{1\over\Lambda^2}\left\{|\lambda|^2\left(|H_1|^2+|H_2|^2\right)
+(\lambda H_1\cdot H_2+{\mathrm h.c.})\right\}
\ .
\ee
With the same superpotential (\ref{Weff}), we can obtain from (\ref{Vgen2LS})
the supersymmetric effective potential for $H_i$ and $t$, which reads:
\bea
V(H_i,t)&=&e^{-2(t+ t^* )/\Lambda}
\left\{{1\over8}(g^2+{g'}^2)\left(|H_1|^2 -|H_2|^2\right)^2
+{g^2\over2}\left(|H_1|^2  |H_2|^2-|H_1\cdot H_2|^2
\right)^2
\right.\nn\\
&+&\left.
|\mu_0|^2
{(|H_1|^2 + |H_2|^2)
\left[1-2(\lambda H_1\cdot H_2+{\mathrm h.c.})/\Lambda^2\right]
+|H_1 \cdot H_2|^2/\Lambda^2
\over
1-|\lambda|^2(|H_1|^2 + |H_2|^2)/\Lambda^2-
(\lambda H_1\cdot H_2+{\mathrm h.c.})/\Lambda^2}
\right\}.
\label{VeffLS}
\eea
This would replace (\ref{Veff}). The $H_i$-dependent part of
this potential, expanded in powers of $H_i/\Lambda$ and keeping
only renormalizable terms reads:
\bea
V(H_i)&= &|\mu_0|^2\left(|H_1|^2 + |H_2|^2\right)+
{|\mu_0|^2\over \Lambda^2}\left|H_1\cdot H_2-
\lambda^*(|H_1|^2 + |H_2|^2)\right|^2\nn\\
&+&{1\over8}(g^2+{g'}^2)\left(|H_1|^2 -|H_2|^2\right)^2
+{1\over2}g^2\left(|H_1|^2  |H_2|^2-|H_1\cdot H_2|^2
\right)^2 ,
\label{VexpLS}
\eea
which would replace (\ref{Vexp}).

The scalar effective potential after inclusion of supersymmetry-breaking 
terms is straightforward to obtain. There is a part quadratic in
$m_{3/2}$:
\be
\delta V_{\rm soft}=m_{3/2}^2\left\{{\Lambda^2\over N_{\rm eff}}
e^{-2(t+t^*)/\Lambda}
-\Lambda^2\left[e^{-3t/\Lambda}+e^{-3t^*/\Lambda}\right]+K_{\rm eff}
\frac{}{}\right\}\ .
\label{softsquaredLS}
\ee
and a part linear in $\tilde{m}_{3/2}$:
\be
\delta V_{\rm soft}=\tilde{m}_{3/2}\left\{
{e^{(t+t^*)/\Lambda}\over N_{\rm eff}}\left[
{\partial W_{\rm eff}\over \partial\varphi_i}
{\partial \Sigma_{\rm eff}\over \partial\varphi_i^*}
-3W_{\rm eff}^*\right]
e^{-3t/\Lambda}
+W_{\rm eff}^* (1-\varsigma)
\frac{}{}\right\}+{\mathrm h.c.}
\label{softlinearLS}
\ee
These equations replace (\ref{softsquared}) and (\ref{softlinear}). 
The SUSY-breaking contribution to the $H_i-t$ effective
potential is long but straightforward to obtain from 
(\ref{softsquaredLS}) and (\ref{softlinearLS}) and we do not give
it explicitly.

Regarding electroweak symmetry breaking, the situation is similar to the one 
presented in section 5. From the minimization conditions and assuming that all
the parameters are real, we obtain again  a first condition for $v$ as
\be
v^2={-m_\beta^2\over \lambda_\beta}\ ,
\label{vB}
\ee
with $m_\beta^2$ as given in (\ref{mbeta}) but now
\be
\lambda_\beta \equiv {1\over 8}(g^2+{g'}^2)\cos^22\beta
+ {1\over 4\Lambda^2}
[-2 \lambda \mu + (\mu-3 \lambda m_{3/2})\sin2\beta]^2\ ,
\label{lbetaB}
\ee
and, a second condition for $\tan\beta$:
\bea
0&=&\cos 2\beta\left\{4 m_{3/2} \Lambda^2\left[
\lambda(2+\varsigma)m_{3/2} - \varsigma \mu
\right]\right.\nn\\
&+&\left.v^2(3 \lambda m_{3/2}  - \mu)\left[2\lambda \mu +
(2\lambda m_{3/2} - \mu)\sin2\beta\right]
\right\}\ ,
\label{tbetaB}
\eea
where, as before,  $\mu\equiv \mu_0 + \lambda m_{3/2}$. In addition we must impose
$m_{\beta}^2<0$ and $\lambda_\beta >0$.
The latter is easily satisfied now, because $\lambda_{\beta} \ge 0$, 
as can be seen from (\ref{lbetaB}); and the former, $m_{\beta}^2<0$, taken together with 
(\ref{tbetaB}), leads to 
\be
\tan \beta = 1.
\ee
The $\tan \beta \neq 1$ solution of eq.(\ref{tbetaB}) is not acceptable for the same 
reasons we found on section~5.

For the Lagrangian of the fermionic sector, we use the general formula 
(\ref{Lfer})
for the SUSY part and (\ref{softfer2}) for the SUSY-breaking piece. After the 
replacement (\ref{repla}), we obtain
\bea
\delta{\cal{L}}_{\rm SUSY}&=&{1\over 2}\left\{
-{\partial^2W_{\rm eff}\over\partial\varphi_i\partial\varphi_j}+
{1\over N_{\rm eff}}
\left[3W_{\rm eff}-{\partial W_{\rm eff}\over\partial\varphi_k} 
{\partial\Sigma_{\rm eff}\over\partial\varphi_k^*}\right]
{1\over\Lambda^2}
{\partial^2 \Sigma_{\rm eff}\over\partial\varphi_i\partial\varphi_j}
\right\}(\chi_i\cdot\chi_j)\nn\\
&+&{2\over\Lambda}{\partial W_{\rm eff}\over\partial\varphi_i}(\chi_i\cdot\chi_t)
-{3W_{\rm eff}\over\Lambda^2}(\chi_t\cdot\chi_t)+{\mathrm h.c.}
\label{LferLS}
\eea
and
\bea
\delta{\cal{L}}_{\rm soft}&=&{1\over 2}{\tilde{m}}_{3/2}e^{-3t/\Lambda}\left\{
-{1\over N_{\rm eff}}{\partial^2 
\Sigma_{\rm eff}\over\partial\varphi_i\partial\varphi_j}(\chi_i\cdot\chi_j)
+2(\chi_t\cdot\chi_t)
\right\}+{\mathrm h.c.}
\label{softferLS}
\eea
To simplify these (and previous)  expressions we have made use of 
$[\partial^2\Sigma_{\rm eff}/\partial
\varphi_i\partial\varphi_j^*]^{-1}=\delta_{ij}$
and $[\partial^3\Sigma_{\rm eff}/\partial
\varphi_i\partial\varphi_j\partial\varphi_l^*]^{-1}=0$.
The conclusions for the neutralino sector are qualitatively similar to 
those of the case discussed in the main text
\newpage

\section*{C. Higgs spectrum}
\setcounter{equation}{0}
\renewcommand{\theequation}{C.\arabic{equation}}

The $3\times 3$ mass matrix ${\bf M_{h}^2}$ for $\cal{CP}$-even Higgses, 
in the basis $\{\phi_i\}=\{h_1^{0r},h_2^{0r},t_R^0\}$, 
with $h_{1,2}^{0r}\equiv \sqrt{2} {\rm Re}[H_{1,2}^0]$ and $t_R^0\equiv \sqrt{2} 
{\rm Re}[t]$
is
\be
\left.
\left[{\bf {M_{h}^2}}\right]_{ij}^{(0)}\equiv 
\left[{\partial^2 V\over\partial \phi_i\partial\phi_j}-\delta_{ij}
{1\over \langle\phi_i\rangle}{\partial V\over\partial\phi_i}\right]
\right|_{\phi_j=\langle\phi_j\rangle}\ ,
\label{Mh}
\ee
where we have subtracted the (zero) tadpoles to get a simpler expression. 
Instead of giving the matrix elements in the basis $\{\phi_i\}$, it proves 
useful to change to the following rotated basis 
$\{\varphi^0\equiv(h_1^{0r}+h_2^{0r})/\sqrt{2}, H^0\equiv(h_1^{0r}-
h_2^{0r})/\sqrt{2}, t_R^0\}$, in which the matrix elements are:
\bea
\langle H^0 | {\bf M_{h}^2}_{(0)} | H^0  \rangle 
&=&2\left[\varsigma\mu m_{3/2}-\lambda(2+\varsigma)m_{3/2}^2
\right]+{1\over 4}(g^2+{g'}^2)v^2\nn\\
&-&v^2{m_{3/2}^2\over\Lambda^2}\left[
\left(\lambda+{13\over 2}\lambda^2\right)
-(2+7\lambda-3\lambda^2){\mu\over m_{3/2}}-
\left(4\lambda+{3\over 2}\right){\mu^2\over m_{3/2}^2}
\right], \nn\\
&& \label{MH}\\
\langle H^0 | {\bf M_{h}^2}_{(0)} | \varphi^0  \rangle &=&0\ ,\\
\langle H^0 | {\bf M_{h}^2}_{(0)} | t_R^0  \rangle &=&0\ ,\\
\langle \varphi^0 | {\bf M_{h}^2}_{(0)} | \varphi^0  \rangle &=&2\lambda_\beta v^2
=\left.v^2{m_{3/2}^2\over\Lambda^2}\right[
\left({13\over 2}\lambda^2
+2\lambda\right)\nn\\
&+&\left.\left(6\lambda^2-{7\over 2}\lambda-4\right){\mu 
\over m_{3/2}}-\left(8\lambda+{7\over 2}\right){\mu^2\over m_{3/2}^2}
\right]\ ,\\
\langle \varphi^0 | {\bf M_{h}^2}_{(0)} | t_R^0  \rangle    
&=&{v\over\sqrt{2}}{m_{3/2}^2\over\Lambda}\left[
-2-(7+3\varsigma)\lambda+(1+3\varsigma){\mu\over m_{3/2}}-
4{\mu^2\over m_{3/2}^2}
\right]\ ,\\
\langle t_R^0 | {\bf M_{h}^2}_{(0)} | t_R^0  \rangle 
&=&m_{3/2}^2+v^2{m_{3/2}^2\over\Lambda^2}\left[
1+{9\over 4}\lambda(3+\varsigma)-{1\over 4}(7+9\varsigma){\mu\over m_{3/2}}+
4{\mu^2\over m_{3/2}^2}
\right]\label{mt}
\ .
\eea
The rest of elements not shown
follow from the symmetric nature of ${\bf M_{h}^2}_{(0)}$. By using this rotated
matrix we have exposed $H^0$ as an exact eigenstate of ${\bf M_{h}^2}_{(0)}$,
with eigenvalue (\ref{MH}). The other two eigenvalues are simply extracted from
the $2\times 2$ submatrix for $\varphi^0-t_R^0$.

The eigenvalues of ${\bf M_{h}^2}_{(0)}$ are not yet the true tree-level 
Higgs masses because a non-minimal K\"ahler potential leads,
after symmetry breaking, to non-canonical kinetic terms for
scalar fields (see the discussion in subsection~3.2). 
More precisely, eq.~(\ref{LK}) gives, for the neutral Higgses
\bea
\delta{\cal{L}}_{\rm kin}&=&\left(\partial_\mu H_1^0 \partial^\mu H_1^{0*}
+\partial_\mu H_2^0 \partial^\mu 
H_2^{0*}
\right)\left[1+(\lambda+1)(\lambda+3){v^2\over 4\Lambda^2}\right]\nn\\
&&+\partial_\mu t \partial^\mu t^*\left[1+(\lambda+1){v^2\over 2\Lambda^2}\right]
-{(1+\lambda)v\over 2\Lambda}\left[ 
\partial_\mu t^* \left( \partial^\mu H_1^0
+\partial^\mu H_2^{0}\right)+{\mathrm h.c.}
\right]\nn\\
&&+(1+\lambda)^2{v^2\over 4\Lambda^2}\left[
 \partial_\mu H_1^0\partial^\mu H_2^{0*}+{\mathrm h.c.}\right]
+{\cal{O}}\left({v^3\over\Lambda^3}\right)\ .
\label{LKnonc}
\eea
Notice that, for the particular value $\lambda=-1$
all  effects associated with non-canonical kinetic terms
disappear. The kinetic terms in (\ref{LKnonc}) 
are recast in canonical form by the field redefinition:
\be
\left[\begin{array}{c}
H_1^0\\
H_2^0\\
t
\end{array}\right]
\rightarrow
{\bf N}
\left[\begin{array}{c}
H_1^0\\
H_2^0\\
t  
\end{array}\right]\ ,
\label{U}
\ee
with ${\bf N}$ a non-singular matrix with positive eigenvalues
of the form ${\bf N=R^{-1}SR}$, where ${\bf R}$ is a field rotation
that makes diagonal the kinetic terms, ${\bf S}$ is a field
re-scaling to get canonical coefficients in the diagonalized 
kinetic terms, and ${\bf R^{-1}}$ rotates the fields back so 
as to obtain ${\bf N=I_3}$ if the kinetic mixing terms in 
(\ref{LKnonc}) were switched off. The explicit form of 
${\bf N}$ is 
\bea
{\bf N}&=&{\bf I_3}+{v\over4\Lambda}(1+\lambda)\left[\begin{array}{ccc}
0 & 0 & 1\\
0 & 0 & 1\\
1 & 1 & 0
\end{array}
\right]\nn\\
&&-{v^2\over 32\Lambda^2}(1+\lambda)
\left[
\begin{array}{ccc}
(9+\lambda) &  (1+\lambda) & 0\\
 (1+\lambda) & (9+\lambda) & 0\\
0 & 0 & 2(1-3\lambda)
\end{array}\right]
+{\cal{O}}\left({v^3\over\Lambda^3}\right)
\ .
\eea
This transformation changes the Higgs mass matrices.
In particular, the new mass matrix for $\cal{CP}$-even
Higgses is
\be
{\bf M_{h}^2}={\bf N}^T {\bf {M_{h}^2}_{(0)}} {\bf N}\ .
\label{UTMU}
\ee
The elements of this new matrix have a simple form in terms of those 
of the initial matrix ${\bf {M_{h}^2}_{(0)}}$ [eqs.~(\ref{MH})-(\ref{mt})].
Note that we are discussing here  an effect up to order 
$v^2/\Lambda^2$, which in principle is small, but 
it is necessary to take it into account 
if we want to know the Higgs masses to order $m^2 v^2/\Lambda^2$ (with 
$m^2=m_{3/2}^2$, $\mu m_{3/2}$, or $\mu^2$). Neglecting contributions of 
order $m^2 v^4/\Lambda^4$ we get
\bea
\langle H^0 | {\bf M_{h}^2} | H^0  \rangle &=&
\langle H^0 | {\bf M_{h}^2}_{(0)} | H^0  \rangle \left[1-
{v^2\over 2\Lambda^2}(1+\lambda)\right]+...\equiv
M_{H^0}^2\nn\\
&=&2\left[\varsigma\mu m_{3/2}-\lambda(2+\varsigma)m_{3/2}^2
\right]+{1\over 4}(g^2+{g'}^2)v^2 \nn\\
&+&v^2{m_{3/2}^2\over\Lambda^2}\left\{
\lambda\left[1-{9\over 2}\lambda+\varsigma(1+\lambda)\right]\right.\nn\\
&-&
\left[\varsigma-2+(\varsigma-7)\lambda+3\lambda^2\right]{\mu\over m_{3/2}}
+\left.{1\over 2}(3+8\lambda){\mu^2\over m_{3/2}^2}\right]\ ,
\label{MHN}\\
\langle H^0 | {\bf M_{h}^2} | \varphi^0  \rangle &=&0\ ,\\
\langle H^0 | {\bf M_{h}^2}| t_R^0  \rangle &=&0\ ,\\
\langle \varphi^0 | {\bf M_{h}^2} | \varphi^0  \rangle &\equiv&
2(\lambda_\beta+\xi_\beta) v^2=
\langle \varphi^0 | {\bf M_{h}^2}_{(0)}| \varphi^0  \rangle \nn\\
&+&{v\over \sqrt{2}\Lambda}(1+\lambda)
\langle \varphi^0 | {\bf M_{h}^2}_{(0)}| t_R^0  \rangle+
{v^2\over 8\Lambda^2}(1+\lambda)^2
\langle t_R^0 | {\bf M_{h}^2}_{(0)}| t_R^0  \rangle+...\nn\\
&=&\left.v^2{m_{3/2}^2\over2\Lambda^2}\right[-{7\over 4}
-{3\over 2}(3+2\varsigma)\lambda
+\left({25\over 4}-3\varsigma\right)\lambda^2\nn\\
&-&\left.\left[7+13\lambda-12\lambda^2-3\varsigma(1+\lambda)\right]
{\mu \over m_{3/2}}-(11+20\lambda){\mu^2\over m_{3/2}^2}\right]\ ,\\
\langle \varphi^0 | {\bf M_{h}^2} | t_R^0  \rangle    
&=&\langle \varphi^0 | {\bf M_{h}^2}_{(0)} | t_R^0  \rangle
+{v\over2\sqrt{2}\Lambda}(1+\lambda)
\langle t_R^0 | {\bf M_{h}^2}_{(0)} | t_R^0  \rangle+...\nn\\
&=&
-{v\over\sqrt{2}}{m_{3/2}^2\over\Lambda}\left[{3\over 2}
+\left({13\over 2}+3\varsigma\right)\lambda-(1+3\varsigma){\mu\over m_{3/2}}+
4{\mu^2\over m_{3/2}^2}\right]\ ,\\
\langle t_R^0 | {\bf M_{h}^2} | t_R^0  \rangle 
&=&\langle t_R^0 | {\bf M_{h}^2}_{(0)} | t_R^0  \rangle 
+{v\over\sqrt{2}\Lambda}(1+\lambda)
\langle \varphi^0 | {\bf M_{h}^2}_{(0)} | t_R^0  \rangle+...\nn\\
&=&m_{3/2}^2+\left.v^2{m_{3/2}^2\over8\Lambda^2}\right[
-1+(20+6\varsigma)\lambda-(25+12\varsigma)\lambda^2\nn\\
&+&\left.2[-5+2\lambda+3(2\lambda-1)\varsigma]{\mu\over m_{3/2}}+
16(1-\lambda){\mu^2\over m_{3/2}^2}
\right]
\ .
\eea
We find that $H^0$ is still an eigenstate of the new mass matrix ${\bf 
M_{h}^2}$ with an eigenvalue, $M_{H^0}$, only slightly different, while 
the $\varphi^0-t_R^0$ submatrix is somewhat changed. 

The $3\times 3$ mass matrix ${\bf M_{A}^2}_{(0)}$ for $\cal{CP}$-odd Higgses,
is obtained by the same formulae (\ref{Mh}) and (\ref{UTMU}),
with $\{\phi_i\}=\{h_1^{0i},h_2^{0i},t_I^0\}$ and 
$h_{1,2}^{0i}\equiv \sqrt{2}{\rm Im}[H_{1,2}^0]$,  $t_I^0\equiv \sqrt{2} 
{\rm Im}[t]$.
In this case it is also useful to perform a change of basis
and work with $\{G^0\equiv (h_1^{0i}-h_2^{0i})/\sqrt{2}, 
a^0\equiv(h_1^{0i}+h_2^{0i})/\sqrt{2}, t_I^0\}$. The neutral Goldstone boson,
$G^0$, is exposed as an eigenstate of this matrix, with zero eigenvalue. 
The remaining $2\times 2$ submatrix has elements
\bea
\langle a^0 |{\bf M_{A}^2}_{(0)}| a^0 \rangle&=&
2\left[\varsigma\mu m_{3/2}-\lambda(2+\varsigma)m_{3/2}^2
\right]\nn\\
&-&v^2{m_{3/2}^2\over\Lambda^2}\left[
\lambda(1+8\lambda)
-(2+10\lambda-3\lambda^2){\mu\over m_{3/2}}-
4\lambda{\mu^2\over m_{3/2}^2}
\right]\ ,\nn\\
&&\label{MAA}\\
\langle a^0 |{\bf M_{A}^2}_{(0)}| t_I^0 \rangle&=&
3{v\over \sqrt{2}}{m_{3/2}^2\over\Lambda}(\varsigma-1)
\left(\lambda-{\mu\over m_{3/2}}\right)\ ,\label{MAt}\\
\langle t_I^0 |{\bf M_{A}^2}_{(0)}| t_I^0 \rangle&=&
9m_{3/2}^2+9v^2{m_{3/2}^2\over\Lambda^2}
(1-\varsigma)\left(\lambda-{\mu\over m_{3/2}}
\right)\ .
\label{Mtt}
\eea
The field redefinition (\ref{U}) affects also the mass matrix for 
pseudoscalars. The elements of the re-scaled matrix ${\bf M_A^2}\equiv
{\bf N}^T{\bf M_{A}^2}_{(0)}{\bf N}$, 
neglecting again terms of order $m^2 v^4/\Lambda^4$,
are\footnote{Since 
${\bf M_{A}^2}_{(0)}$
is written in the basis $\{G^0,a^0,t_I^0\}$, 
eqs.~(\ref{MAA})-(\ref{Mtt}),
then ${\bf N}$ in $(\ref{U})$ has to be rotated to that same basis.}
\bea
\langle a^0 |{\bf M_{A}^2}| a^0 \rangle&=&
\langle a^0 |{\bf M_{A}^2}_{(0)}| a^0 \rangle\left[1-{v^2\over 4\Lambda^2}
(1+\lambda)(5+\lambda)\right]\nn\\
&+&{v\over\sqrt{2}\Lambda}(1+\lambda)\langle a^0 |{\bf M_{A}^2}_{(0)}| t_I^0 \rangle
+ {v^2\over 8\Lambda^2}(1+\lambda)^2\langle t_I^0 |{\bf M_{A}^2}_{(0)}| t_I^0 \rangle+
...\nn\\
&=&
2\left[\varsigma\mu m_{3/2}-\lambda(2+\varsigma)m_{3/2}^2
\right]\nn\\
&+&\left.v^2{m_{3/2}^2\over8\Lambda^2}\right\{9+
(18+22\varsigma)\lambda
+(24\varsigma-43)\lambda^2
+2(2+\varsigma)\lambda^3\nn\\
&-&2\left.\left[
(11+12\lambda+\lambda^2)\varsigma-2(7+23\lambda-6\lambda^2)\right]
{\mu\over m_{3/2}}+32\lambda{\mu^2\over m_{3/2}^2}
\right\}\ ,\nn\\
&&\\
\langle a^0 |{\bf M_{A}^2}| t_I^0 \rangle&=&
\langle a^0 |{\bf M_{A}^2}_{(0)}| t_I^0 \rangle+
{v\over2\sqrt{2}\Lambda}(1+\lambda){\mathrm Tr}\ {\bf M_{A}^2}_{(0)}\nn\\
&=&{v\over\sqrt{2}}{m_{3/2}^2\over\Lambda}\left\{{9\over 2}
+\left(2\varsigma-{1\over 2}\right)\lambda
-(2+\varsigma)\lambda^2+\left[
3+(\lambda-2)\varsigma\right]
{\mu\over m_{3/2}}\right\}
\ ,\nn\\
&&\\
\langle t_I^0 |{\bf M_{A}^2}| t_I^0 \rangle&=&
\langle t_I^0 |{\bf M_{A}^2}_{(0)}| t_I^0 \rangle\left[1-{v^2\over 4\Lambda^2}
(1+\lambda)(1-3\lambda)\right]\nn\\
&+&{v\over\sqrt{2}\Lambda}(1+\lambda)\langle a^0 |{\bf M_{A}^2}_{(0)}| t_I^0 \rangle
+ {v^2\over 8\Lambda^2}(1+\lambda)^2\langle t_I^0 |{\bf M_{A}^2}_{(0)}| t_I^0 \rangle+
...\nn\\
&=&
9m_{3/2}^2+\left.v^2{m_{3/2}^2\over8\Lambda^2}\right\{-9-4(2\varsigma-5)\lambda+
(7+8\varsigma)\lambda^2\nn\\
&-&\left.2(2+\varsigma)\lambda^3+2\left[
-3+(\lambda-2)^2\varsigma+6\lambda\right]
{\mu\over m_{3/2}}\right\}
\ .
\eea
This shows that, for $\mu$ and $m_{3/2}$ moderately larger than $v$, $a^0$ 
and $t_I^0$ are approximate mass eigenstates
up to ${\cal{O}}(v/\Lambda)$ corrections. Their masses can be read 
off directly from the formulae above.

The $2\times 2$ mass matrix ${\bf{M_{ch}^2}}_{(0)}$ for charged Higgses, 
in the basis $\{\phi_i^+\}=\{H_1^{+},H_2^+\}$, 
is obtained from a formula similar to (\ref{Mh}). 
By a suitable change of basis, the charged Goldstone boson 
$G^+\equiv (H_1^{+}-H_2^+)/\sqrt{2}$ can be exposed as an eigenstate
of this matrix, with zero eigenvalue. The other eigenstate
is the physical Higgs $H^+\equiv (H_1^{+}+H_2^+)/\sqrt{2}$ 
with mass given by the trace of ${\bf{M_{ch}^2}}_{(0)}$. 
Using (\ref{MH}), it can be related to the mass of $H^0$ as
\be
\langle H^- |{\bf{M_{ch}^2}}_{(0)}| H^+ \rangle
=\langle H^0 |{\bf{M_{h}^2}}_{(0)}| H^0 \rangle
-{1\over 4}{g'}^2v^2\ .
\ee

Once again, to get the true mass we need to take into account that kinetic 
terms
should be canonical. Up to ${\cal{O}}(v^2/\Lambda^2)$, the kinetic 
piece of the Lagrangian for charged Higgses reads
\be
\delta{\cal{L}}_{\rm kin}=\left[\partial_\mu H_1^- \left(\partial^\mu H_1^{-}\right)^*
+\partial_\mu H_2^+ \left(\partial^\mu H_2^{+}\right)^*\right]
\left[1+{v^2\over 2\Lambda^2}(1+\lambda)\right]\ .
\ee
This is canonical again after the re-scaling
\be
\left[\begin{array}{c}
H_1^+\\
H_2^+
\end{array}\right]
\rightarrow
{\bf N'}
\left[\begin{array}{c}
H_1^+\\
H_2^+
\end{array}\right]\ ,
\label{Up}
\ee
with
\be
{\bf N'}\equiv \left[1-{v^2\over 4\Lambda^2}(1+\lambda)\right]{\bf I_2}\ .
\ee
The corrected mass for the charged Higgs boson is therefore
\be
M_{H^+}^2\equiv\langle H^- |{\bf{M_{ch}}^2}_{(0)}| H^+ \rangle
\left[1-{v^2\over 2\Lambda^2}(1+\lambda)\right]
=M_{H^0}^2-{1\over 4}{g'}^2v^2\ ,
\ee
with $M_{H^0}^2$ as defined in eq.~(\ref{MHN}).


\end{document}